\documentclass[fleqn,usenatbib]{mnras}

\usepackage{newtxtext,newtxmath}
\usepackage{subcaption}

\usepackage[T1]{fontenc}
\usepackage{comment}
\usepackage[normalem]{ulem}
\DeclareRobustCommand{\VAN}[3]{#2}
\let\VANthebibliography\thebibliography
\def\thebibliography{\DeclareRobustCommand{\VAN}[3]{##3}\VANthebibliography}

\usepackage{graphicx}	
\usepackage{amsmath}	
\usepackage{url}

\usepackage{color,ulem}  
\definecolor{deepblue}{rgb}{0,0,0.5}  
\definecolor{deepred}{rgb}{0.6,0,0}   
\definecolor{deepgreen}{rgb}{0,0.5,0} 
\definecolor{darkgreen}{rgb}{0,0.6,0} 

\newcommand{\AN}[1]{{{\bf{\color{magenta}AN--[}}{\color {darkgreen}#1]}}} 

\def\cmed{{$c_{(1/2)}$}\/}

\def\civ{{C\sc{iv}$\lambda$1549}\/}

\def\D{$\Delta$\/}

\def\ergs{erg s$^{-1}$}
\def\feii{{Fe\sc{ii}}\/}

\def\fcha{F$_{\rm C,H\alpha}$}
\def\fbha{F$_{\rm B,H\alpha}$}
\def\frha{F$_{\rm R,H\alpha}$}
\def\fchb{F$_{\rm C,H\beta}$}
\def\fhb{F$_{\rm H\beta}$}
\def\gsoft{$\Gamma_{\rm soft}$}
\def\ha{{\sc{H}}$\alpha$\/}
\def\habc{{\sc{H}}$\alpha_{\rm BC}$}

\def\hb{{\sc{H}}$\beta$\/}
\def\hbbc{{\sc{H}}$\beta_{\rm BC}$}

\def\he{{\sc{H}}$\epsilon$\/}

\def\kms{km~s$^{-1}$\/}
\def\l{$\lambda$}

\def\lbol{L$_{\rm bol}$\/}

\def\ledd{$L_{\rm Edd}$\/}

\def\l5100{L$_{\rm 5100}$}
\def\nh{n$_{\rm{H}}$}

\def\mbh{M$_{\rm BH}$}
\def\mgii{Mg{\sc ii}$\lambda$2800}

\def\oiii{{[O\sc{iii}]}\/}
\def\oiiia{{[O\sc{iii}]}$\lambda$4959\/}
\def\oiiib{{[O\sc{iii}]}$\lambda$5007\/}
\def\oiiill{{[O\sc{iii}]}$\lambda\lambda$4959,5007\/}

\def\redd{R$_{\rm{Edd}}$\/}
\def\rfe{R$_{\rm{FeII}}$\/}

\def\rk{R$_{\rm K}$}
\def\rfbha{R$_{\rm FB,H\alpha}$}
\def\rfrha{R$_{\rm FR,H\alpha}$}
\def\rhb{R$_{\rm FH\beta}$}

\title[Index Diagram to disentangle AGN-HG]{SDSS-V Black Hole Mapper: The Index Diagram as a tool to disentangle the influence of the Host Galaxy in Quasar spectra 
}

\author[C. A. Negrete et al.]{
C. A. Negrete,$^{1,2}$\thanks{E-mail: alenka@astro.unam.mx}
R. Sandoval-Orozco,$^{1}$
H. Ibarra-Medel,$^{1}$
B. Tapia,$^{3}$ 
R. J. Assef,$^{5}$
D. Dultzin,$^{1}$ 
\and
I. Lacerna,$^{8,9}$
S. Morrison,$^{11}$
S. F. Anderson,$^{18}$
P. Rodr\'iguez Hidalgo$^{24}$
C. Aydar,$^{4,14}$
F. E. Bauer,$^{15}$
\and
E. Benitez,$^{1}$
D. Bizyaev,$^{21,22}$
W. N. Brandt,$^{6,7,12}$
J. R. Brownstein,$^{20}$
J. Buchner,$^{4}$
I. Cruz-Gonz\'alez,$^{1}$
\and
D. Gonz\'alez-Buitrago,$^{1,2}$
H. Hern\'andez-Toledo,$^{1}$
N. Jenaro-Ballesteros,$^{1}$
A. Koekemoer,$^{13}$
Y. Krongold,$^{1}$
\and
M. L. Mart\'inez-Aldama,$^{25}$
K. Pan,$^{21}$
C. Ricci,$^{5,10}$
M. Salvato,$^{4}$
S. F. S\'anchez,$^{1,23}$
D. Serrano-F\'elix,$^{1}$
\and
D. P. Schneider,$^{6,7}$
M. Sniegowska,$^{17}$
B. Trakhtenbrot,$^{4,17}$
Q. Wu,$^{11}$ 
D. Wylezalek,$^{16}$
Q. Yang,$^{19}$
\and
R. J. Zermeño,$^{1}$
\\
$^{1}$Universidad Nacional Autónoma de México. Instituto de Astronomía. A.P. 70-264, 04510. Ciudad de México, México.\\
$^{2}$SECIHTI Research Fellow\\
$^{3}$Instituto de F\'isica - Universidad Nacional Aut\'onoma de M\'exico\\
$^{4}$Max-Planck Institute for Extraterrestrial Physics, Giessenbachstrasse1, 85748, Garching, Germany\\
$^{5}$ Instituto de Estudios Astrof\'isicos, Facultad de Ingenier\'ia y Ciencias, Universidad Diego Portales, Av. Ej\'ercito Libertador 441, Santiago, Chile\\
$^{6}$Department of Astronomy \& Astrophysics, The Pennsylvania State University, University Park, PA 16802, USA\\
$^{7}$Institute for Gravitation and the Cosmos, The Pennsylvania State University, University Park, PA 16802, USA\\
$^{8}$Instituto de Astronom\'ia y Ciencias Planetarias, Universidad de Atacama, Copayapu 485, Copiap\'o, Chile\\
$^{9}$Millennium Institute of Astrophysics (MAS), Nuncio Monseñor S\'otero  Sanz 100, Providencia, Santiago, Chile\\
$^{10}$Kavli Institute for Astronomy and Astrophysics, Peking University, Beijing 100871, People's Republic of China\\
$^{11}$Department of Astronomy, University of Illinois at Urbana-Champaign, Urbana, IL 61801, USA\\
$^{12}$Department of Physics, The Pennsylvania State University, University Park, PA 16802, USA\\
$^{13}$Space Telescope Science Institute, 3700 San Martin Dr., Baltimore, MD 21218, USA\\
$^{14}$Excellence Cluster ORIGINS, Boltzmannstrasse 2, D-85748 Garching, Germany\\
$^{15}$Instituto de Alta Investigaci\'on, Universidad de Tarapac\'a, Casilla 7D, Arica, Chile\\
$^{16}$Astronomisches Rechen-Institut, Zentrum fur Astronomie der Universitat Heidelberg, Monchhofstr. 12-14, D-69120 Heidelberg, Germany\\
$^{17}$School of Physics and Astronomy, Tel Aviv University, Tel Aviv 69978, Israel\\
$^{18}$Department of Astronomy, University of Washington, Box 351580, Seattle, WA 98195, USA\\
$^{19}$Center for Astrophysics, Harvard \& Smithsonian, 60 Garden St., Cambridge, MA 02138, USA\\
$^{20}$Department of Physics and Astronomy, University of Utah, 115 S. 1400 E., Salt Lake City, UT 84112, USA\\
$^{21}$Apache Point Observatory and New Mexico State University, Sunspot, NM 88349, USA\\ 
$^{22}$Sternberg Astronomical Institute, Moscow State University, 119234, Moscow,
Russia\\ 
$^{23}$Instituto de Astrof\'\i sica de Canarias, La Laguna, Tenerife, E-38200, Spain\\
$^{24}$ Physical Sciences Division, University of Washington Bothell, WA, 98011, USA\\
$^{25}$ Universidad de Concepci\'on, Chile\\
}

\date{Accepted XXX. Received YYY; in original form ZZZ}

\pubyear{2015}

\begin{document}
\label{firstpage}
\pagerange{\pageref{firstpage}--\pageref{lastpage}}
\maketitle

\begin{abstract}
We revisit the Quasar Main Sequence (QMS) by investigating the impact of the stellar component from the host galaxy (HG) on the emission line spectra of the active galactic nuclei (AGN). 
We first detect spectra with broad emission lines using a line ratio method for a sample of $\sim$3000 high SNR ($>$20) Black Hole Mapper objects (part of the fifth phase of the Sloan Digital Sky Survey).
We then built the Index diagram, a novel diagnostic tool using the $z$-corrected spectra, model-free, designed to easily identify spectra with significant stellar HG contributions and to classify the AGN spectra into three categories based on AGN-HG dominance: HG-dominated (HGD), Intermediate (INT), and AGN-dominated (AGND) sources. A colour-$z$ diagram was used to refine the AGN-HG classification. 
We subtract the stellar contributions from the HGD and INT spectra before modeling the AGN spectrum to extract the QMS parameters.
Our QMS reveals that HGD galaxies predominantly occupy the Population B region with no \rfe, 
with outliers exhibiting \rfe\ $>$ 1, likely due to HG subtraction residuals and a faint contribution of \hbbc. 
INT and AGND spectra show similar distributions in the Population A 
region, while in Population B, 
a tail of AGND sources becomes apparent. 
Cross-matching with radio, infrared, and X-ray catalogs, we find that the strongest radio emitters are associated with HGD and INT groups. Strong X-ray emitters are found in INT and AGND sources, also occupying the AGN region in the WISE colour diagram.

\end{abstract}




\defcitealias{CortesSuarez2022}{CS22}
\defcitealias{borosongreen92}{BG92}
\section{Introduction}
\label{sec:Intro}


Type 1 AGN are those objects that show broad lines in their optical-UV spectra and are considered unobscured objects. Since the geometry of the central region is not spherical, the observational properties, particularly the line profiles from the broad line region (BLR), depend on the orientation of the accretion disc (AD) with respect to our line of sight (LOS). The accretion rate is a second factor that determines the composition and shape of the broad emitting lines. Other physical properties such as density, ionization parameter, metallicity, and dynamics of the BLR also influence the observed line intensity and profiles. 
All these combined result in a spectral zoo that encompasses not only the optical-UV region but the entire spectral energy distribution (SED). 

One of the most useful and powerful tools to organize and classify in an elegant way the large spectral diversity of Type 1 AGN and quasars is the so-called Quasar Main Sequence (QMS). 
The QMS originates from
the principal component analysis from which Eigenvector 1 (E1) was derived \citep[][]{borosongreen92,sulentic00a, shenho14, Marziani2022}. 
An E1 in 4 dimensions (4DE1) has been proposed \citep[e.g.][]{50yearsOfQuasars}, covering two optical dimensions, one in the UV and one in X-rays: (1) the FWHM(\hbbc), (2) the ratio of \feii\ intensities in the optical (between 4435 and 4685 \AA) and \hbbc, \rfe\ (3) the mid-height centroid of \civ, \cmed\ and (4) the index in soft X-rays \gsoft. The selection of these four dimensions is based on the representativeness of the central region components. The FWHM is related to the kinematics of the BLR; the \feii\ is associated with the AD emission (both \hb\ and \feii\ are low ionization lines), the \cmed\ measures the wind component in \civ, a high ionization line, and the \gsoft\ is associated with the X-ray corona, where the ionizing photons come from. 

Considering the optical range of the QMS, two populations are defined in terms of the FWHM(\hbbc) \citep[see e.g., Figure 1 of][]{zamfir10}. Population A is defined as having FWHM < 4000 \kms, where mostly radio-quiet objects (RQ) are found, lower SMBH (log\mbh\ $\sim$ 6-8.5 to $z$ up to 0.8), and a wide \redd\ range. 
Since the range of \rfe\ is wide, subpopulations A1, A2, A3, A4 are defined in terms of increments in \D\rfe\ = 0.5, finding a trend toward A3-A4 with \redd\ close to or greater than 1, high densities (log\nh\ $\sim$ 12-14), low ionization parameter (logU $\sim$ -3 -- -4) and high metallicity \citep[usually super solar][]{negrete12, Garnica2022, Buendia-Rios2023}. The reason why these objects can reach regimes close to or super Eddington is that as the accretion rate increases, the disc becomes thick in an advection dominant accretion flow (ADAF) regime so that the luminosity of the source tends to saturate \citep[][and references there in]{M18}. Therefore, due to this effect it is possible to use these objects 
as cosmological candles \citep{MS14, Negrete2017, Sandoval-Orozco2024, Buendia-Rios2025}. Regarding the line profile, it has been found that a single Lorentzian is the best-fitted profile. On the other hand, Population B objects have the opposite characteristics of Population A, with FWHM > 4000 \kms, most objects are radio-loud (RL), the SMBH tend to be more massive (log\mbh\ $\sim$ 7.5-9.8 at $z$ up to 0.8), the \redd\ range is lower than 0.2, with low densities and high ionization parameters \citep{negrete13,negrete14}. The B1, B1+, and B1++ subpopulations are defined in terms of increments in \D FWHM = 4000 \kms, with some objects falling into population B2 (\rfe\ between 0.5 and 1 and FWHM between 4000-8000 \kms).

The optical plane of the QMS (FWHM(\hb) vs. \rfe) has a delimited distribution of points that define a sequence of spectral types and permits us to separate into sub-populations as described above. Sources belonging to the same spectral type show similar spectroscopic line profiles, line flux ratios, and other spectral characteristics, which permit us to assume similar broad line physics and geometry. 
Thus, the 4DE1 sequence has been defined by following a main sequence (hence the term Quasar Main Sequence) expressed in terms of the increase of the mass of the supermassive black hole (SMBH, \mbh) towards Populations Bs, anticorrelated to the Eddington ratio \redd\ = \lbol/\ledd\ (the ratio of the bolometric luminosity \lbol, and the Eddington luminosity \ledd\ = 1.26 $\times 10^{38}$ M/M$_\odot$ \ergs), also considered a dimensionless accretion rate \citep{marziani01}, towards Populations A3-A4.

The luminosity distribution in AGN at low $z$ (up to 0.8) is wide, with ranges of log\lbol\ = 43-47. However, due to a selection effect, the least luminous objects are found at $z$ below 0.5. 
Additionally, due to the SMBH and stellar mass \mbh-M* relation, in the low \redd\ regime (towards 0.001), the luminosity of the host galaxy (HG) can compete with the luminosity of the nuclear region, so we can observe the stellar continuum mixed with the broad and narrow emission lines from the active nucleus. The contamination of the stellar component mainly affects the continuum measurement, biasing the measurements to values above the true value. 
Also, stellar absorption features can be confused with \feii\ multiplet emission around \hb\ \citep{bon20}. 
At the other extreme, where \redd\ tends to have values close to 1, the stellar contribution is overshadowed by nuclear emission. \citet{negrete18} showed for a sample of $\sim$300 spectra from the SDSS DR7 (Sloan Digital Sky Survey Data Release) that automatic measurements of AGN spectra in the optical without considering the stellar continuum lead to erroneous estimates of the QMS optical plane parameters (even with high S/N > 20), affecting their subsequent analysis. In a later work, \citet{bon20} analysed the spectra of the \citet{negrete18} sample with host galaxy features by decoupling the stellar component from the AGN emission. The result was that they showed that objects initially classified as population A3-A4 actually belonged to population B1-B2-B1+, contaminating the sample of cosmological candles.

Other works using hundreds of thousands of spectra using different SDSS data releases have also reconstructed and analysed the QMS. For instance, \citet{WuShen2022} analysed $\sim$750k spectra from the SDSS DR17, reconstructing a QMS with large scatter. They explain the high dispersion by the fact that the mean S/N of the catalogue is $\sim$5, and therefore, they consider that more sophisticated model fits that include the stellar component are not possible and will return inaccurate results for the majority of low-z quasars. 

A first work to organize AGN spectra with stellar HG contamination was done by \citet[][hereafter \citetalias{CortesSuarez2022}]{CortesSuarez2022}, using the MaNGA survey\footnote{Mapping Nearby Galaxies at Apache point observatory is the fourth phase of the SDSS (https://www.sdss4.org/surveys/manga/.}, which is a survey dominated by galaxies without active nuclei. After identifying Type 1 AGN spectra, they used indices around \hb\ and the CaII H band to separate three kinds of spectra according to the AGN-HG dominance.  

AGN-HG decoupling has been challenging due to the degeneracy between the stellar continuum and the AGN power law \citep[e.g.][]{VandenBerk2006}. Works such as \citet{Husemann2013,Husemann2014} and  \citet{Ibarra-Medel2025} have developed 
tools to separate the AGN-HG emission by taking advantage of integral field spectroscopy, considering each spaxel as a set of independent spectra. Basically, the stellar contribution is considered using a 3D galaxy model as input and iterating between the HG emission and the nuclear region emission from the broad line flux. For two-dimensional spectra, it is common to use multiple component decomposition codes such as {\sc PyQSOFiT} \citep{Guo2018PyQSOFit} or a modification of \textit{Starlight} \citep{CidFernandes2005}, which includes spectral modelling of the observed spectra using single stellar population and multicomponent emission line models and power-law templates to disentangle the HG spectra form the AGN emission. 

Therefore, it is crucial, and is the main goal of this work, to have a method that allows us to detect spectra of AGN and quasars with stellar contribution, while at the same time detecting the "contamination level" of the HG to the AGN power law. Having spectra free of HG emission will allow us to better determine the properties of nuclear active galaxies using existing diagnostic diagrams, such as QMS. 
This paper is organized as follows. Section \ref{sec:sample} reports the sample selection method. Section \ref{sec:methods} describes the spectral characterization of our selected sample. Section \ref{sec:stl_pyqsofit} refers to the methodology used for spectral analysis. In Section \ref{sec:results}, we present the results of our data analysis, while in Section \ref{sec:discussion} we discuss the implications of our findings in a multiwavelength context. Finally, in Section \ref{sec:summary}, we summarize our results.

\section{Sample description}
\label{sec:sample}
\begin{figure}
    \includegraphics[width=\columnwidth]{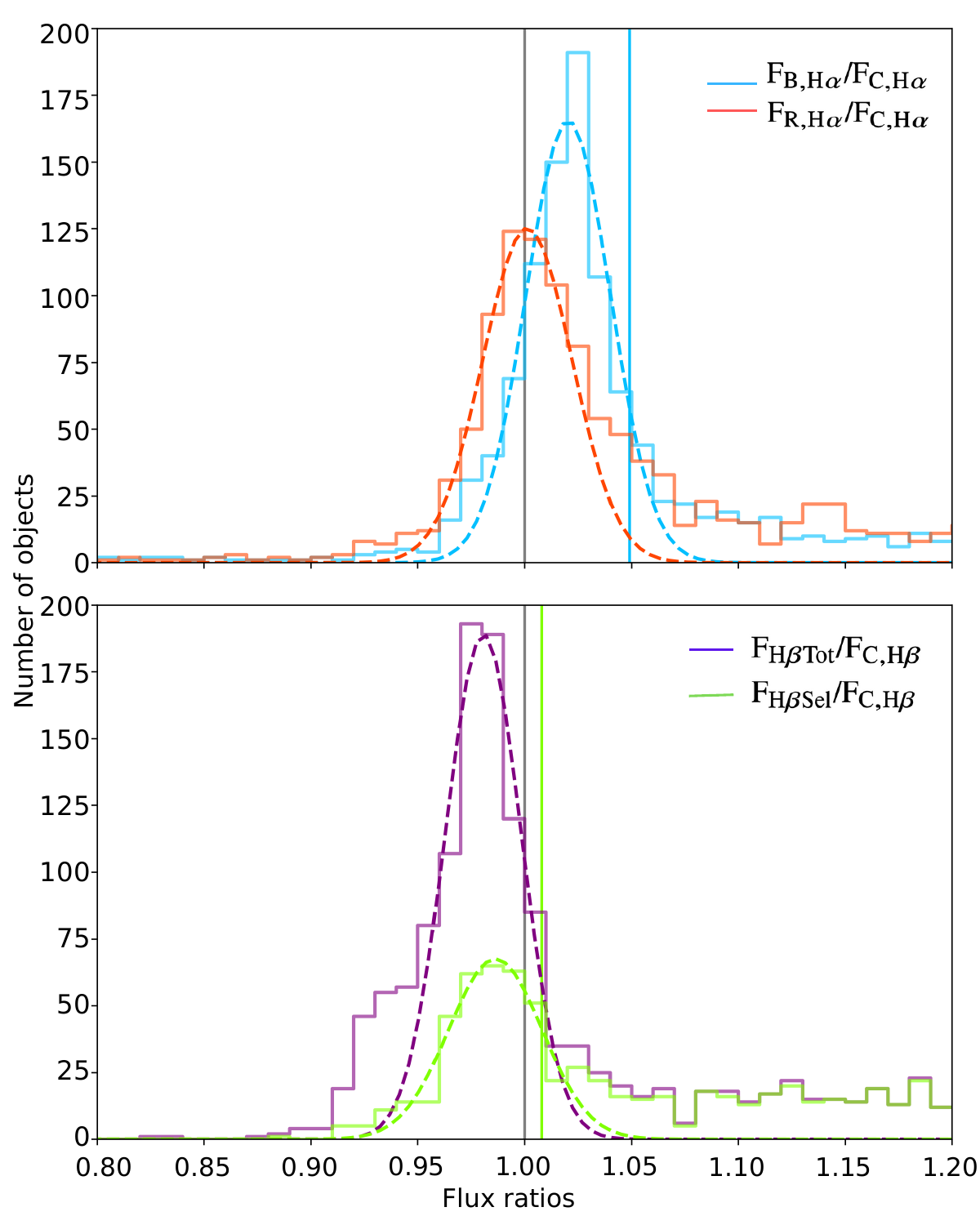}
    \caption{Zoom in flux ratio distributions for objects with $z \leq$ 0.5. Solid lines are the distributions in bins of 0.001. Dashed lines are the fitted Gaussians (see text for details). Vertical black lines indicate a flux ratio equal to one, while blue and green vertical lines are the limits for F$_{\rm B,H\alpha}$/F$_{\rm C,H\alpha}$ and F$_{\rm H\beta Sel}$/F$_{\rm C,H\beta}$, respectively (see Table \ref{tab:histograms}). \textit{Top panel}: Blue lines: F$_{\rm B,H\alpha}$/F$_{\rm C,H\alpha}$. Orange lines: F$_{\rm R,H\alpha}$/F$_{\rm C,H\alpha}$. \textit{Lower Panel}: Purple lines: F$_{\rm H\beta Tot}$/F$_{\rm C,H\beta}$. Green lines: F$_{\rm H\beta Sel}$/F$_{\rm C,H\beta}$. The vertical axis is the number of objects in each bin.}
    \label{fig:histograms}
\end{figure}

We use spectroscopical data from the Sloan Digital Sky Survey phase V survey (SDSS-V). The SDSS-V is the first all-sky multi-epoch and optical/infrared spectroscopic survey \citep[Kollmeier et al. in prep][]{Almeida+2023}. It maps the two hemispheres simultaneously at the 2.5-m telescope at the Apache Point Observatory (APO) in the northern hemisphere and the 100-inch du Pont Telescope at Las Campanas Observatory (LCO) in the southern hemisphere. Both telescopes have the modified Multi-object fibre-fed spectrograph of the Baryon Oscillation Spectroscopic Survey \citep[BOSS][]{Smee+2013}, and the Apache Point Observatory Galactic Evolution Experiment (APOGEE) spectrograph \citep{Wilson+2019}. The BOSS spectrograph obtains optical spectral data with 500 fibres with a spectral resolution of $R\approx$ 2000 and a spectral range of 3600 to 10,000 \AA. The APOGEE spectrograph obtains IR spectral data with 300 fibres across the H-Band (1.51-1.70 $\mu m$) wavelength range with a spectral resolution of $R\approx$ 22,000. In addition to the two spectrographs, each telescope uses the new robotic fibre positioner named the Focal Plane System (FPS, Medan et al. in prep), which defines the new multi-object spectroscopy (MOS) system. This MOS can reconfigure at real-time 500 targets within an overhead of 3 minutes and across a field of view of 2 degrees. This MOS obtains the Optical/IR multi-epoch spectra of two of the three mappers that form the SDSS-V: The Black Hole Mapper (BHM) and the Milky Way Mapper. In addition to this infrastructure, a new observatory was constructed at LCO, implementing a new and innovative Integral Field Unit (IFU) with 1801 lenslet fibers in a 0.5-degree diameter. 
This IFU obtains the optical spectral and spatially resolved data for the last SDSS-V mapper: The Local Volume Mapper \citep[][]{Drory2024}. This work will use the BOSS optical multi-epoch spectral data from the BHM. The scientific goal of the BHM is to provide a physical understanding of the process surrounding the growth and evolution of supermassive black holes (SMBH) and their coevolution with other host galaxies. 
The BHM consists of three main core programs: The Reverberation mapping (RM) program, which plans to obtain multiepoch spectra for more than 1,500 targets with cadences as short as 3 days for a total of 100-150 epochs per target. 
The All-Quasar Multi-Epoch Spectroscopy (AQMES) program plans to observe 20,000 targets with a modest cadence spanning 2 to 10 epochs. And the SPectroscopic IDentification of eROSITA Sources (SPIDERS), which plans to obtain a follow-up optical spectroscopical survey of the X-ray eROSITA survey and plans to observe at least 300,000 targets.

\subsection{Sample selection}
\label{sec:sample_tot}

We used the SDSS BHM data observed up to December 2021, which consists of 177,977 spectra. 
The first selection criterion is selecting spectra with positive redshift values, considering the one given in the header. 
We also limit our sample to only those targets that have $g-r$ observed colours between -1 and 3. Those colour values need to be at the observed frame, without any k-correction applied in order to consider the colour reddening of the hosts due to the redshift displacement. With these two criteria, we cleaned the sample to obtain 143,125 spectra. 

For each object, we made a coadded spectrum corrected by the Galactic extinction. We also set the restframe wavelength using the redshift given by the SDSS-V. We obtained a total of 52,030 spectra for individual objects within a redshift range of 0.015 $< z <$ 4.511, each one having between one and 64 spectra. 
Each spectrum has an exposure time of 900 seconds, so the number of coadded spectra equals exposure times between 900 and 57,600 seconds.

High signal-to-noise (S/N) spectra are needed to accurately decompose the emission line components, so we choose objects with S/N $\geq$ 20 in the continuum. For each spectrum, we compute the S/N in the r band and the continuum windows around 1350, 1700, 2200, 3000, 5100, and 6200 \AA, to select objects with at least one of those S/N values larger than 20. We also considered the median S/N reported by the survey. Within this criterion we find 10,269 objects (20\% of the initial sample). For this work, we are interested in the region around \hb\ to build the E1 optical plane, so we choose objects with $z \leq$ 0.9 and find 3259 spectra, which we call the complete sample (CoSa).

Spectra at $z \leq$ 0.5 are generally less luminous than those at higher z, and the contribution of the host galaxy is more likely to be detected. Moreover, \ha\ can only be observed at $z \leq$ 0.5. Additionally, we need to be sure that our sample is only composed of broad-line AGN, so we decided to split the sample of 3,259 spectra into high and low $z$ ranges to make a second selection. We have 2,092 spectra in the $z$ range 0--0.5 
and 1,167 objects at $z$ 0.5--0.9. 

\subsubsection{Sample at $z \leq$ 0.5}
\label{sec:sample_0-05}

\begin{table}
	\centering
	\caption{Values of the flux ratio distributions of Figure \ref{fig:histograms} for objects with $z \leq$ 0.5. For each range, we report the mode ($\mu$), the standard deviation ($\sigma$), the upper limits (U$_{\rm L}$), and the number of objects above each limit with the percentage that represents. Bold numbers are the Upper limits considered for the selection (see text for details).}
	\label{tab:histograms}
	\begin{tabular}{lccccc} 
		\hline
		Range & $\mu$ & $\sigma$ & U$_{\rm L}$ & Num Obj & Percentage\\
		\hline
		\ha B   & 1.027 & 0.022 & 1.049 & 1337 & 64\%\\
		\ha R   & 1.003 & 0.018 & 1.021 & 1581 & 76\%\\
		\hb & 0.980 & 0.018 & 0.998 & 1374 & 66\%\\
		\hb$_{\rm sel}$ & 0.986 & 0.022 & 1.008 & 1134 & 54\%\\
		\hline
	\end{tabular}
\end{table}

To detect and discard low-z ($z \leq$ 0.5) spectra that do not have broad emission lines, 
we follow a methodology similar to the one described in \citetalias{CortesSuarez2022} (see also \citealt{Oh2015}). The method identifies a broad component in emission for \ha\ (\habc) based on the flux ratio between a band around \ha\ (on the blue or red side, \fbha\ and \frha) and a band around a region in the continuum (\fcha), far enough so that \habc\ does not contaminate it. In this way, Type 1 AGN can be identified up to Type 1.9, i.e., those objects that only show a broad \ha\ line but not in \hb\ \citep{Osterbrock1981}. 
The regions are 6520-6540 \AA, 6590-6610 \AA, and 6400-6420 \AA\ for the blue band, red band, and continuum, respectively (see Figure 1 on \citetalias{CortesSuarez2022}). 
In contrast to the MaNGA objects, our sample is dominated by AGN, so we used an additional approach after the band ratio calculations to detect the spectra without \habc. 

We built histograms of the line ratio distributions of \rfbha\ = \fbha/\fcha\ and \rfrha\ = \frha/\fcha\ shown in blue and red solid lines in Figure \ref{fig:histograms}, using bins of 0.01 in flux ratio. 
We fit a Gaussian to the data distribution given by values lower than the maximum of the distribution (the mode, $\mu$) plus one bin of values greater than $\mu$ (to avoid the long tail distribution towards larger values of the flux ratios, up to $\sim$5) to estimate its standard deviation $\sigma$ (blue and red dashed lines in Figure \ref{fig:histograms}, which is a zoom in of the line ratio distributions). 
The $\mu$ and $\sigma$ values of each distribution are 1.027 and 0.022 for \rfbha, and 1.003 and 0.018 for \rfrha\ (see Table \ref{tab:histograms}). In this way, we will consider that objects having values larger than the upper limit U$_{\rm L} = \mu+\sigma$ in each blue and red band have a broad component in \ha. The number of objects with flux ratio values above U$_{\rm L,BH\alpha}$=1.049 and U$_{\rm L,RH\alpha}$=1.021 are 1,322 and 1,577, respectively. Our limits are slightly lower in comparison with \citetalias{CortesSuarez2022}, who computed U$_{\rm L}$ values of 1.056 and 1.037 for the blue and red bands. 
We note that \citetalias{CortesSuarez2022} use a boxplot-whiskers diagram where the upper limit was computed using the quartiles of the distribution. However, the distributions of the two samples are different: more than 90\% of the MaNGA galaxies are quiescent, while in our sample, the vast majority is expected to have an active nucleus. Therefore, we decided not to use the boxplot-whisker method for our analysis.

Unlike \citetalias{CortesSuarez2022}, who considered both red and blue flux ratios to restrict the candidates to detect \habc, we will only consider the blue flux ratio because our sample is contaminated by the skylines at the red end of the spectra close to $z \sim$ 0.4-0.5. We find 1,321 spectra above U$_{\rm L,BH\alpha}$, which represent 63\% of the objects with $z \leq$ 0.5. A visual inspection of the rest of the objects (772) revealed that 16 had Balmer broad component, of which five had wrong $z$ assignment (two of them show only the \hb\ range), another nine have small absorptions or sky subtraction residuals, and the last two have a very broad component that goes beyond F$_{\rm C,H\alpha}$. Adding these 16 objects, we have 1,337 spectra above U$_{\rm L,BH\alpha}$ that are our candidates for having \habc. 

For our study, to build the E1 optical parameter space, it is necessary to have the presence of the broad component of \hb\ (\hbbc). So we follow a flux ratio procedure similar to the one for the detection of \habc\ now considering bands around \hb. We will use only a flux band at the blue side of \hb, around 4830-4850 \AA\ (\fhb), since the red side is contaminated by both \oiiia\ and \feii\ Opt42 emissions. The \hb\ continuum window (\fchb) was taken at 4950-4970 \AA. Following the same procedure as in \ha, we find that the peak distribution of the flux ratios \rhb\ = \fhb/\fchb\ is $\mu_{\rm H\beta}$ = 0.980 with a $\sigma$ = 0.018. The solid purple line in Figure \ref{fig:histograms} illustrates the former distribution, while the dashed purple line is the Gaussian distribution fitted with the same criteria as in \ha. Having a $\mu_{\rm H\beta}$ close but lower than one means that we have a considerable tail in the distribution towards objects with no \hbbc. 
If we fit a Gaussian restricting the bin numbers between minus two and plus three bins around $\mu_{\rm H\beta}$, we obtain an upper limit U$_{\rm H\beta,tot}$ = 0.998. This upper value means that 34\% of the objects at $z \leq$ 0.5 do not have \hbbc. 
If we consider a distribution of objects with \rfbha\ above the U$_{\rm L,BH\alpha}$=1.049 (solid green line in Fig. \ref{fig:histograms}), we retrieve values of $\mu_{\rm H\beta,sel}$ = 0.986. Fitting a Gaussian in the same way as for \rfbha\ (dashed green line) we obtain a $\sigma_{\rm H\beta,sel}$ = 0.022, and a U$_{\rm H\beta,sel}$ = 1.008. 
Above this last upper limit, we find 1,126 objects. Table \ref{tab:histograms} summarizes the values for the four flux ratio distributions described above.

Finally, we performed a visual inspection of the rejected objects and identified eight spectra with wrong redshift assignment: three showing \hb\ emission, another three with small absorptions in the flux bands, and the last two with extremely broad components that reach the flux continuum range. So, the number of candidates for having \hbbc\ is 1,134 objects at $z \leq$ 0.5.

\subsubsection{Sample at 0.5 $< z \leq$ 0.9}
\label{sec:sample_05-09}

As we move towards higher z, most objects have bluer colours than their low-z counterpart, so we expect a larger contribution of broad-line AGN spectra. So, we did not perform line ratio analysis as for the low-z sub-sample. In contrast, the remaining 1167 objects at 0.5 $< z \leq$ 0.9 were inspected visually to detect deviations such as wrong $z$ values or sky residuals. We found 54 objects with sky residuals in the red end of the spectra, but the \hb\ region was sufficiently clear to make the line decomposition. Another 13 spectra have wrong flux calibration, and 81 have wrong $z$ determination, of which 43 are optical and 38 are UV spectra. Removing from the sample the wrong flux-calibrated spectra and the UV spectra, we have 1116 good-quality spectra. 

\subsection{Redshift Correction}
\label{sec:z-corr}

\begin{figure}
    \centering
    \includegraphics[width=\columnwidth]{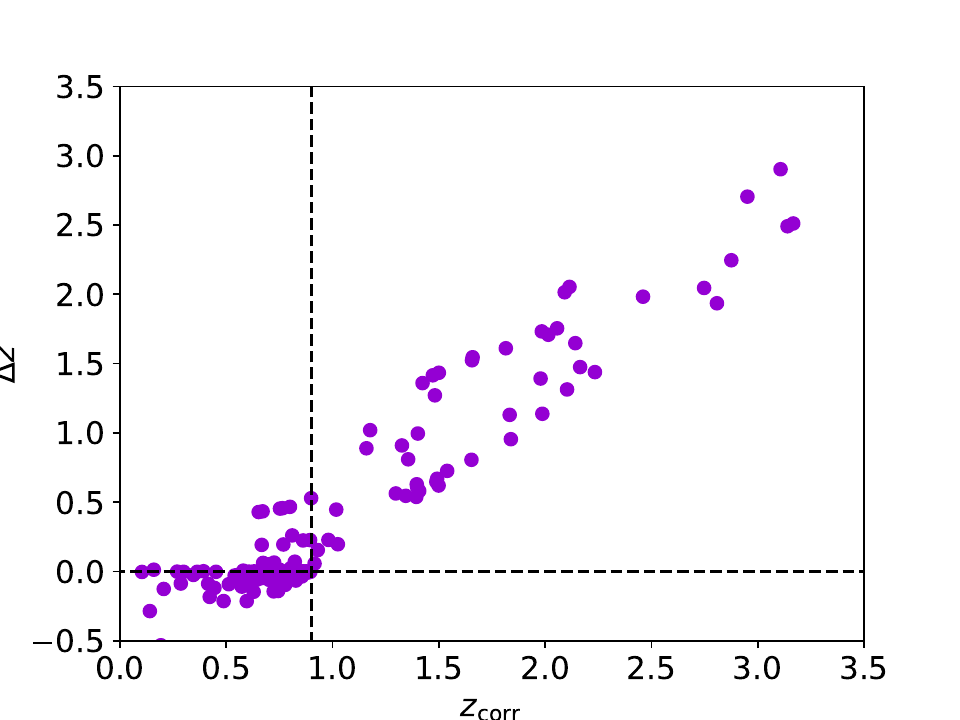}
    \includegraphics[width=\columnwidth]{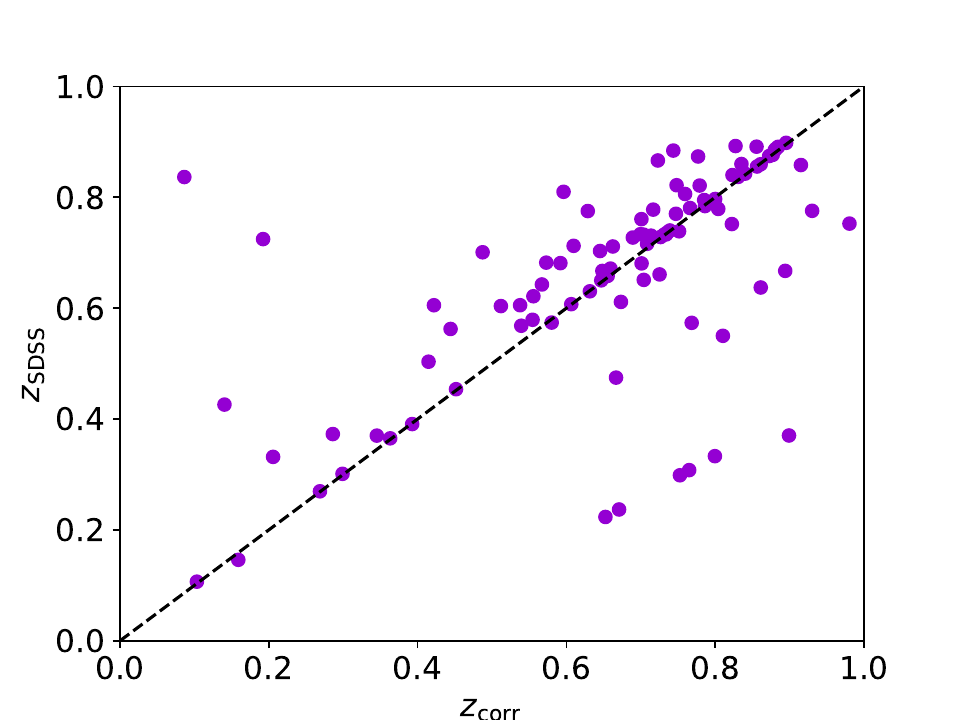}
    \caption{
    $\Delta z = z_{SDSS} - z_{corr}$ (upper panel) and $z_{SDSS}$ for objects with $z <$ 1 (lower panel) as functions of the corrected redshift distribution ($z_{corr}$) for objects with wrong $z$ assignment. 
    }
    \label{fig:zcorr}
\end{figure}

It has been reported that redshift computations in the SDSS ($z_{SDSS}$) have some deviations \citep{HewettWild2010, Bolton2012, negrete18}, showing a disagreement in the rest-frame wavelength of the spectral lines when corrected by the provided $z$. By visual inspection, we found 143 objects with wrong $z$ computation. Of them, 54 have good-quality optical spectra and were corrected by recomputing the redshift ($z_{corr}$), measuring the strongest visible lines i.e., \mgii, \hb, and \oiiib. The range of the $\Delta z$ corrections lies between -0.750 and 0.467 ($\Delta z/(1+z)$ from -0.690 to 0.259). 
Another 41 spectra of the low-z region show small shifts between -0.034 and 0.014 in $z$. 
We considered it necessary to correct for $z$ shifts of less than $\pm$ 0.01 because in the PyQSOfit fits (Sec. \ref{sec:PyQsofit}), the \oiii\ and \hb\ NCs were shifted with respect to the rest frame. A $\Delta z$ = 0.01 implies a shift of $\sim$3000 \kms. We also corrected objects showing UV quasar spectra 
with redshifts up to $z \sim$ 3.167. 
Figure \ref{fig:zcorr} shows the $z_{corr}$ distribution 
against the difference of $\Delta z = z_{SDSS} - z_{corr}$ along 0.1 $> z >$ 3.2 (upper panel), and $z_{SDSS}$ for objects with $z <$ 1 (lower panel). 

\subsection{Sample composition}
\label{sec:sample-composition}

We started considering an initial sample (CoSa) of 3259 objects with a S/N $\geq$ 20 and a $z$ range up to 0.9. We have to reject spectra without emission lines (12\%), 
flat spectra without emission lines and objects with severe sky residual contamination (4\%), 
objects showing only narrow emission lines like starburst galaxies or Type 2 AGN (14\%), 
which we did not analyse, as they are outside of the scope of this paper, and objects with wrong $z$ assignment (1.4\%). 


Our final sample (FiSa) consists of 2336 broad-line AGN, 51.6\% 
at $z \leq$ 0.5 and 48.4\% 
in the $z$ range between 0.9 and 0.5. For our FiSa, 4.6\% of the objects are Type 1.9 AGN. 

\section{AGN spectral Characterization}
\label{sec:methods}

\begin{figure}
    \includegraphics[scale=0.54]{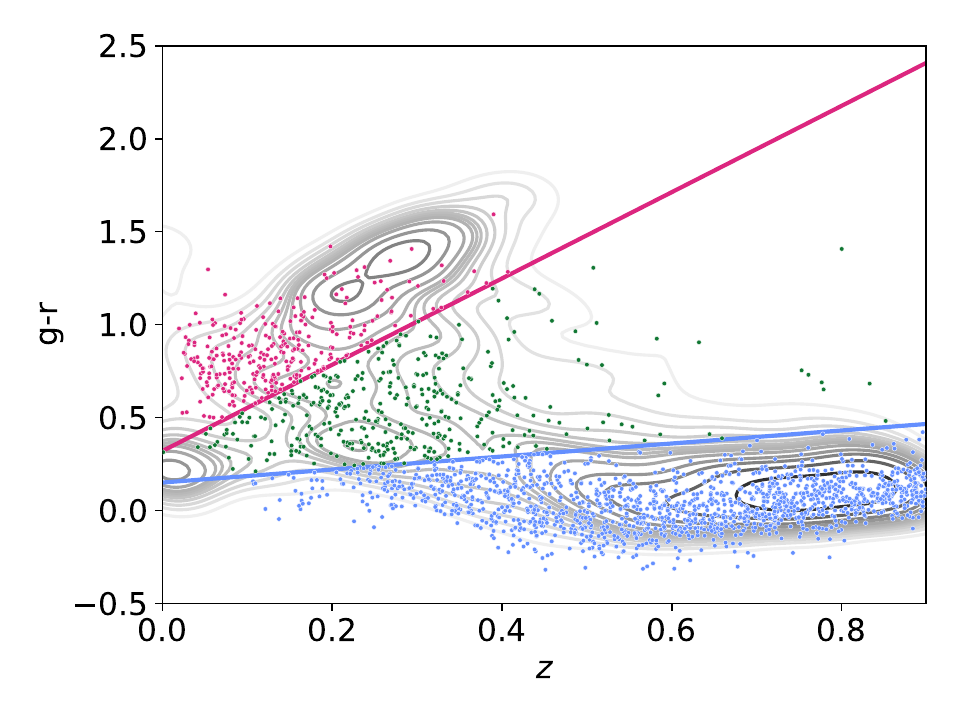}\\
    \includegraphics[scale=0.27]{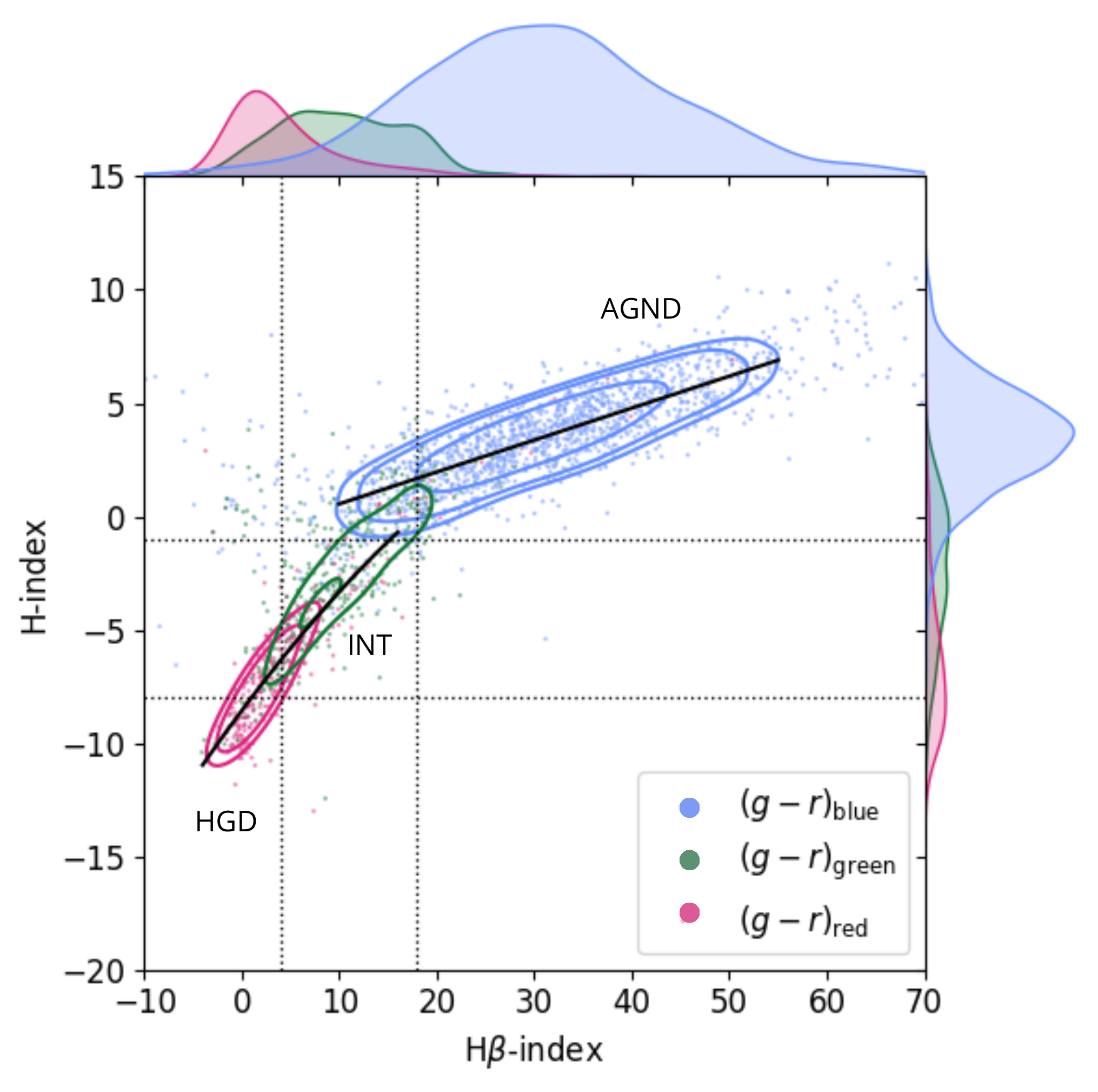}
    \caption{\textit{Upper panel}: Colour--$z$ diagram using the CoSa. The ordinate axis is the redshift, while the abscissa is the $g-r$ colour of the PSF aperture in magnitudes. Solid red line represents the lower limit of the red spectra. Solid blue line is the upper limit of the bluer spectra. Red, Green, and Blue dots are the corresponding objects of each group.  
    Over-plotted black isocontours are meant to enhance the density distribution of a sample without a S/N restriction (see text for details). 
    \textit{Lower panel}: H-\hb\ index diagram using the CoSa. Dashed black vertical and horizontal lines delimit the AGND, HGD, and INT groups described in the text. The colour points, their isocontours, and the histograms correspond to the points delimited by $g-r$ in the figure of the upper panel. Solid black lines are the correlations given in eqs. \ref{eq:index_AGND} and \ref{eq:index_HGD-INT} for AGND and HGD+INT groups, respectively.
    }
    \label{fig:color-index}
\end{figure}

In Section \ref{sec:Intro} we discussed that the origin of the AGN spectral diversity is governed principally by the accretion rate and the orientation of the accretion disc with respect to our line-of-sight. For low-$z$ low luminosity AGN, the contribution of the stellar continuum of the HG to the nuclear emission should also be considered. 
As spectral differences can be large, it is not possible to analyse all the AGN spectra in a similar way. In this section, we propose two methods (diagnostic diagrams) to separate the quasar spectra into groups according to the AGN intensity with respect to the HG contribution. 

We will use the CoSa 
(which includes spectra with no emission lines, strong sky subtraction residuals and Sy2 galaxies) with the purpose of showing that the method is useful for a mixed sample. In a forthcoming paper, we will prove our diagnostic classification (described in this section) with a more homogeneous sample, including, for example starburst galaxy spectra and Sy2 galaxies. We only remove objects with z$_{corr}$ larger than 0.9.


\subsection{Colour-z diagnostic diagram}
\label{sec:ColorDiagram}

The colour-$z$ diagram is a straightforward relation presented in the upper panel of Fig. \ref{fig:color-index}. This colour-$z$ diagram uses the SDSS observed $g-r$ Gun AB colours. The colour-$z$ plot shows that the object distribution follows two main groups or families, with a sparse third group in the middle. Looking at the distribution of all the points in the redshift range $z <$ 0.5, we can see a mixture of colours, including $g-r$ values around zero or even negatives, which are indicators for flat or negative slopes. We identify a first family of objects at $z <$ 0.5 having an increasing colour value, i.e., these are reddened objects (Red zone; red points in Fig. \ref{fig:color-index}, with $g-r \geq$ 0.5) where the contribution of the HG becomes important. In those spectra, the cosmological shift from their rest frames generates a reddening effect from the observed $g-r$ colour: the K-correction. Therefore, the observed colour is mainly accentuated by the shift of the Balmer break through the photometric bands, with a $z$-dependent slope from $z$=0 towards $z$=0.5. In general, at $z <$ 0.5 it is possible to find both HG- or AGN-dominated spectra 
(see Sec. \ref{sec:IndexDiagram} for a description of the HG and AGN dominance), although our low-$z$ sample is dominated by HG-dominated objects.  

The second group is largely composed of spectra with bluer colours corresponding to luminous AGN: the large majority are located at $z >$ 0.5 (Blue zone; blue points in Fig. \ref{fig:color-index}), showing colour in the range -0.3 $\leq g-r \leq$ 0.3. These spectra have negative slopes originating from the power law of the active nucleus of the form $f_\nu \sim \nu^{-\alpha}$, i.e., the dominant contribution comes from the non-thermal emission. We also observe a third group in the transition of those two regions with colours $g-r$ from 0 to 1.5 (Green zone; green points in Fig. \ref{fig:color-index}). 

We over-plot isocontours of the spectra with $z < 0.9$ without a S/N restriction (18,867 objects) to enhance the density distribution that clusters in the regions delimited by objects above the line that follows the equation 
\begin{equation}
    (g-r)_{\rm red} = 2.3 \, z + 0.32 
\end{equation}
that corresponds to 42\% of our CoSa, and below the delimitation line %
\begin{equation}
    (g-r)_{\rm blue} = 0.35 \, z + 0.15
\end{equation}  
having 27\% of our CoSa in this region. The remaining objects that compose the 31\% are in the transition region that we defined as $(g-r)_{\rm green}$. In a forthcoming paper (Ibarra-Medel et al. in prep.), we will explore the stellar contribution and population for the complete quasar sample, including Sy2 objects.

\subsection{The {\rm H}-\hb\ Index diagram}
\label{sec:IndexDiagram}

The second diagnostic diagram is the H-\hb\ Index diagram presented in \citetalias{CortesSuarez2022}. They used indices in two regions: the CaII H band (the H-Index) in the 3950-3990 \AA\ region and the Lick \hb\ index (the \hb -Index) defined in the 4848-4878 \AA\ spectral interval in the rest frame. Our indices are mathematically defined as follows: 
\begin{equation}
    \text{Index}=\int_{\lambda_0}^{\lambda_1}\frac{f(\lambda)\,-<\text{Cont}>}{<\text{Cont}>}d\lambda,
\end{equation} 
where $\lambda_0$ and $\lambda_1$ are the upper and lower limits of the index region, $f(\lambda)$ is the observed spectral flux density in erg/s/cm$^2$/\AA, and $<Cont>$ are the average flux of the adjacent continuum of each index. 
The indices are computed using a continuum window with a width of 50\AA\ around 4020 and 5100 \AA, for CaII H and \hb\ respectively, taking into account the slope of the Balmer jump for the H-Index. 
We followed the \citetalias{CortesSuarez2022} terminology, where positive values are for emission lines, and negative values are for absorption lines. 
\citetalias{CortesSuarez2022} choose those indices because when the AGN spectra are dominated by the host galaxy's stellar contribution (Host Galaxy dominated objects, HGD), the CaII H band is strong, and the \hb\ emission line could be ``hidden'' (absorbed) by the HG. As the AGN power increases, the CaII H band is replaced by the non-thermal continuum plus the 
\he\ emission line that emerges from the nuclear region, \he\ being the most important contribution to the H-Index when the AGN dominates. Following this line, the AGN-dominated galaxies (AGND) should have both indices (\hb- and H-Index) in emission. There is also an intermediate AGN group (INT) where the \hb\ broad component is seen, and at the same time, we can see a clear host galaxy contribution. 

In the lower panel of Figure \ref{fig:color-index}, we show the distribution of the CoSa 
in the H-\hb\ index diagram. We over-plot the group separations reported in \citetalias{CortesSuarez2022}, shown in dashed vertical lines of \hb-index at 4.0 and 18.0, and dashed horizontal lines of H-index at -1.0 and -8.0. to classify our objects into
\begin{itemize}
    \item HGD with \hb-Index $<$ 4.0 and H-Index $<$ -1
    \item INT with \hb-Index $>$ 4.0 and H-Index $<$ -1, and
    \item AGND with H-Index $>$ -1
\end{itemize}

The isocontours are intended to emphasize the distribution of the three $g-r$ colour groups in the index diagram. 
The histograms of each $g-r$ colour group are presented in the upper and right panels to enhance their distributions along the H-\hb-index diagram. Colour code is according to the regions delimited in the colour-$z$ diagram. As expected, HGD galaxies have redder colours, AGND objects are bluer, and intermediate objects are in the colour transition. 

In the H-\hb\ index diagram, we observe two different trends: one for the AGND objects that follow the equation of a straight line
\begin{equation}
\label{eq:index_AGND}
    (H-\mathrm{index})_{\rm AGND} = 0.0774* H\beta-\mathrm{index} + 1.0995,
\end{equation}
and a second trend that originates from the HGD sample towards the Intermediate objects following the second-order equation
\begin{equation}
\label{eq:index_HGD-INT}
\begin{split}
    (H-\mathrm{index})_{\rm HGD,INT} = & -0.0032 (H\beta-\mathrm{index})^2 \\ 
    & + 0.4127 (H\beta-\mathrm{index}) - 6.0315.
\end{split}
\end{equation}

The main difference between the trends described above and those found by \citetalias{CortesSuarez2022} is the sample size and the $z$ range. They used 47 objects at $z \leq$ 0.15 (delimited by the MaNGA survey), with which the bimodality of the distribution of the AGND and HGD+INT objects cannot be observed. In both CoSa and FiSa, the large majority of objects (60\% and 79\% respectively) have $z > 0.5$. 

From now on, we will use red colour code for objects in the HGD region, green for those in the INT region, and blue for those in the AGND region.

\begin{figure}
    \centering
    \begin{subfigure}{0.5\textwidth}
        \includegraphics[width=1\linewidth]{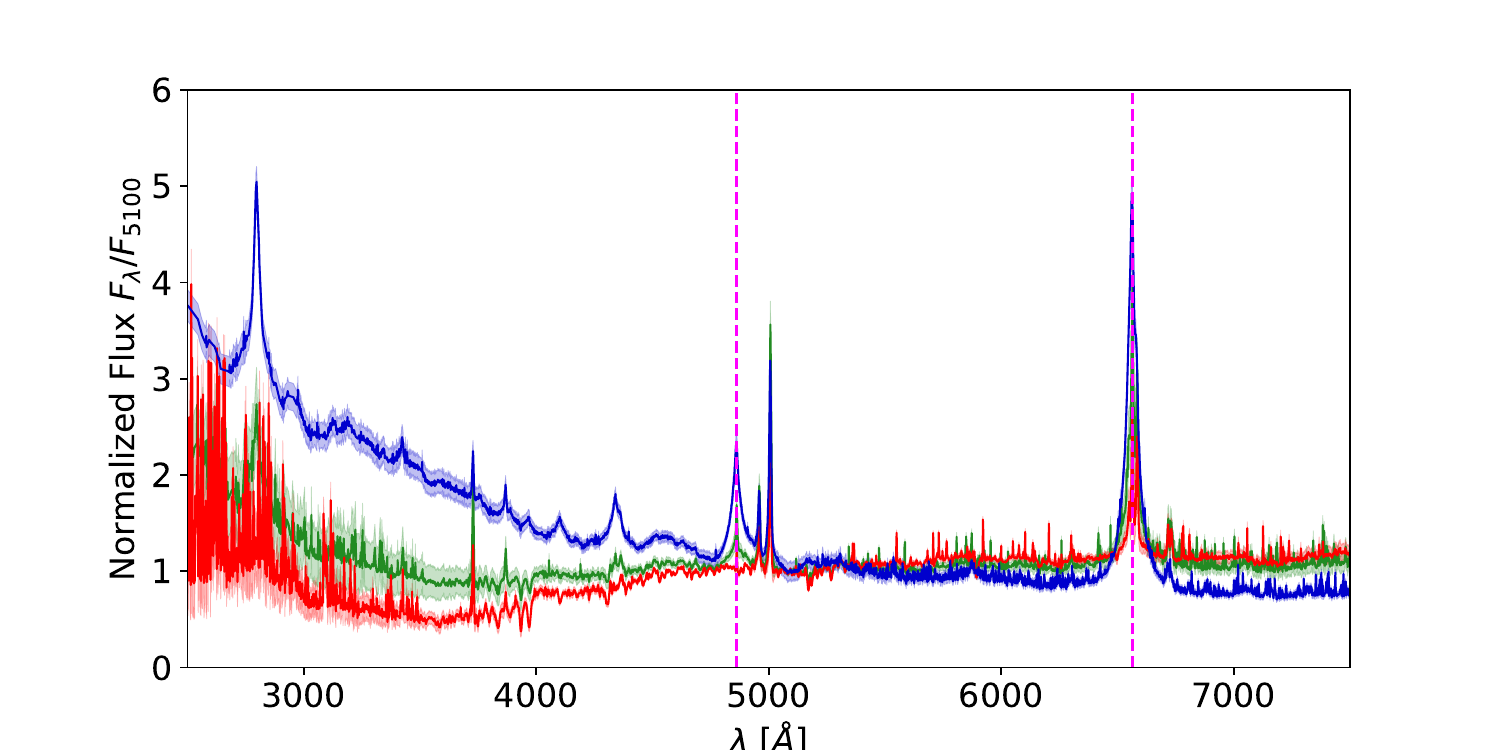}
        \caption{Average spectra of the selected objects.}
    \end{subfigure} \vfill
    \begin{subfigure}{0.5\textwidth}
        \includegraphics[width=1\linewidth]{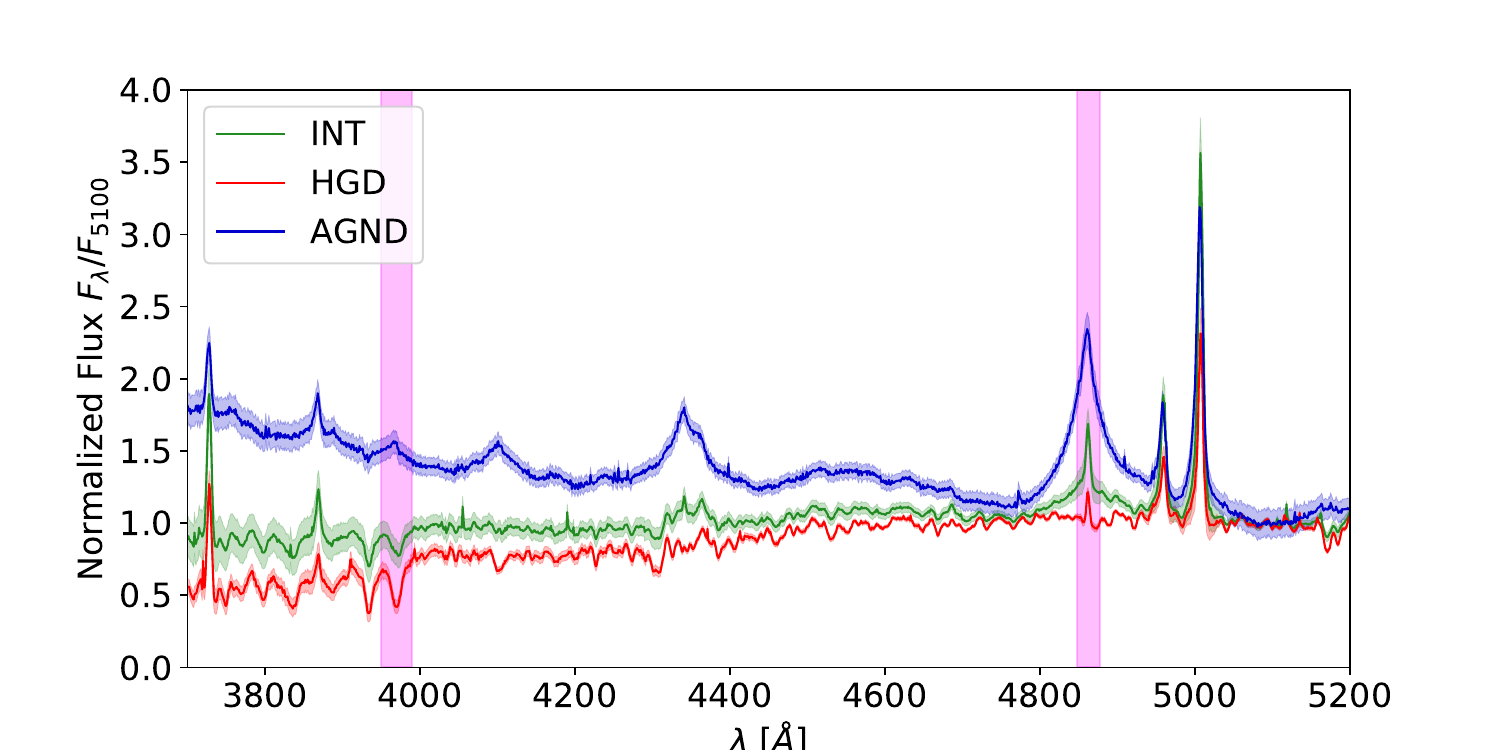}
        \caption{Zoom in the region between 3900-5200 \AA.}
    \end{subfigure} \vfill
    \caption{(\textit{Upper panel}) Average spectra of the selected groups. The shaded region is the variance interval, following the colour code for each spectrum, where blue represents AGND, red represents HGD, and green represents INT sources. Vertical dashed lines are the rest frames for \ha\ and \hb. (\textit{Lower panel}) Zoom-in plot to visualize the H and \hb\ indices regions shown in pink bands.}
    \label{fig:avg_spectra}
\end{figure}

Figure \ref{fig:avg_spectra} illustrates the coadded spectra of the three HGD, INT and AGND groups. The upper panel shows the full visible range between 2400-7500 \AA, with dashed pink vertical lines indicating the \ha\ and \hb\ rest frames. The lower panel shows a zoom-in between 3700-5200 \AA, where the H and \hb\ indices spectral regions are illustrated within the pink vertical regions.

The distribution of the CoSa objects in the H-\hb\ index diagram is: 27\% are in the HGD region, 14\% in the INT, and 59\% in the AGND region.
Objects with no emission lines (neither broad nor narrow) and narrow emission line objects, 
have red $g-r$ colours and lie in the HGD zone.
Regarding the distribution of the FiSa in the H-\hb\ index diagram, we find 13\%, 17\%, and 70\% percentages for the HGD, INT, and AGND groups, respectively. 

\section{Stellar subtraction and AGN modeling}
\label{sec:stl_pyqsofit}
Once we have delimited our groups for the FiSa according to the AGN-HG dominance, 
we proceed to obtain the spectral characterization according to the QMS. 
In particular, we are interested in the region around \hb\ but we will also characterize the \ha\ region as well (for objects at $z < 0.5$). 
We decided to keep the spectra dominated by the HG with no visible \hb\ broad component, initially classified as Sy 1.9, because in some cases this component may become significant once the stellar continuum is subtracted, shifting its classification to an earlier Seyfert type \citep[e.g.][]{bon20}. 

\subsection{Stellar Continuum subtraction}

As we described in Sec \ref{sec:IndexDiagram}, 30\% of the FiSa objects are in the HGD or INT groups. Spectra with HG contribution need to be separated from the AGN emission before the spectral analysis. For this purpose, we use the \textit{Starlight} code \citep[STL,][]{CidFernandes2005} to disentangle both emissions. For the stellar population fitting, we first mask the emission line contribution by considering a large width window for the permitted lines, specially of the Balmer lines and Fe, with a greater width for the INT spectra.  
In the INT spectra, we also assume a contribution from the \feii\ underlying pseudo continuum around \hb. We also added to the STL a set of ten power laws templates with slopes from 0 to -3 to fit the non-thermal continuum originating from the AGN continuum. For the AGND, we initially assume that the AGND spectra do not show a significant HG contribution. However, as discussed below, in some AGND spectra, a faint HG contribution could be seen and could affect the measurement of the power-law continuum flux. For AGND spectra, it is hard to separate the stellar contribution using STL; the mask needs to consider all the permitted broad lines along with the strong FeII emission. In consequence, the stellar fitting regions are small and few in number. So we decided not to use STL decomposition for the AGND group. For AGND spectra with a weak stellar contribution (7.7\% of all AGND), the maximum error in the continuum flux measurements is $\sim$5\%. In those spectra, the errors in the \rfe\ and FWHM(\hbbc) computations are negligible.

\begin{figure}
    \centering
    \includegraphics[scale=0.52]{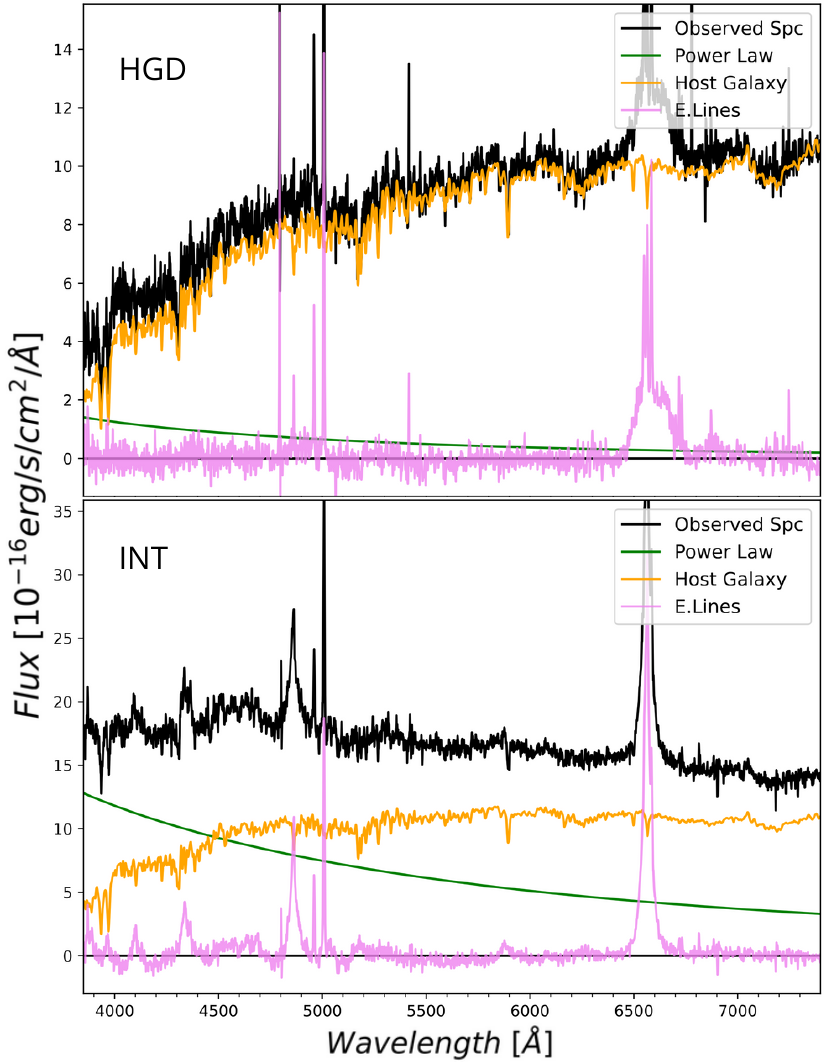}
    \caption{Examples of stellar-AGN decomposition for HGD (left) and INT (right) groups. The solid black line is the observed spectrum, the yellow line is the stellar contribution from the HG, the green line is the AGN power law, and the pink line is the AGN emission line spectrum.}
    \label{fig:STL}
\end{figure}

Figure \ref{fig:STL} shows two examples of stellar-AGN decomposition for HGD (upper panel) and INT (lower panel) spectra. The solid black line is the observed spectrum, the yellow line is the stellar contribution from the HG, the green line is the AGN power law, and the pink line is the ``cleaned'' AGN emission line spectrum. We will consider the AGN spectra as the sum of the power law (PL) plus the emission line spectrum (AGN+PL) 
and not only the one of the AGN, because STL gives a hint of the PL. As expected, we see a faint AGN contribution in the HGD sample spectra in the \hb\ region, but \ha\ broad component is clearly seen. The AGN+PL contribution in the INT spectra is stronger, and in consequence, the stellar absorption continuum is less prominent than in the HGD case. 

As the decomposition of the PL and the stellar continuum could be degenerate \citep[e.g.][]{Ibarra-Medel2025}, we computed the error associated with the fitted models by 
measuring the variance of 500 simulated spectra by adding a random noise contribution to each spectral type, HGD or INT. 
Those spectra were analysed using the same STL algorithm to calculate the error of the continuum with the PL and the Stellar contribution. Using the STL outputs for the HGD objects, 
we computed an associated error of $\sim 0.11$ for the PL in flux units ($\times 10^{-16} erg/s/cm^{2}$/\AA), 
while the Stellar contribution is associated with an error of $0.89$ in flux units. 
Meanwhile, the INT selected objects have errors of $\sim 0.14$ and $\sim 0.86$ in flux units for the PL and the Stellar contribution, respectively. This represents an error of $\sim 20\%$ of the total flux. As mentioned above, the degeneracy between the PL and the stellar continuum decoupling implies, in cases for fitting a larger PL, the need for a lesser contribution of the stellar continuum and viceversa.

The effect of the HG subtraction on the QMS has been considered by several previous studies \citep[e.g.][]{Zamfir2008,negrete18,bon20}. For example, \citet{Zamfir2008} find a small change of the FWHM(\hb) of about 5\%\ when the quasar emission was cleaned from the stellar continuum, while the \feii\ emission was affected with a 25–35 per cent, spanning a wide range of the QMS populations. 

\subsection{Spectral Fitting}

\label{sec:PyQsofit}
\begin{figure*}
    \centering
    \includegraphics[width=0.8\linewidth]{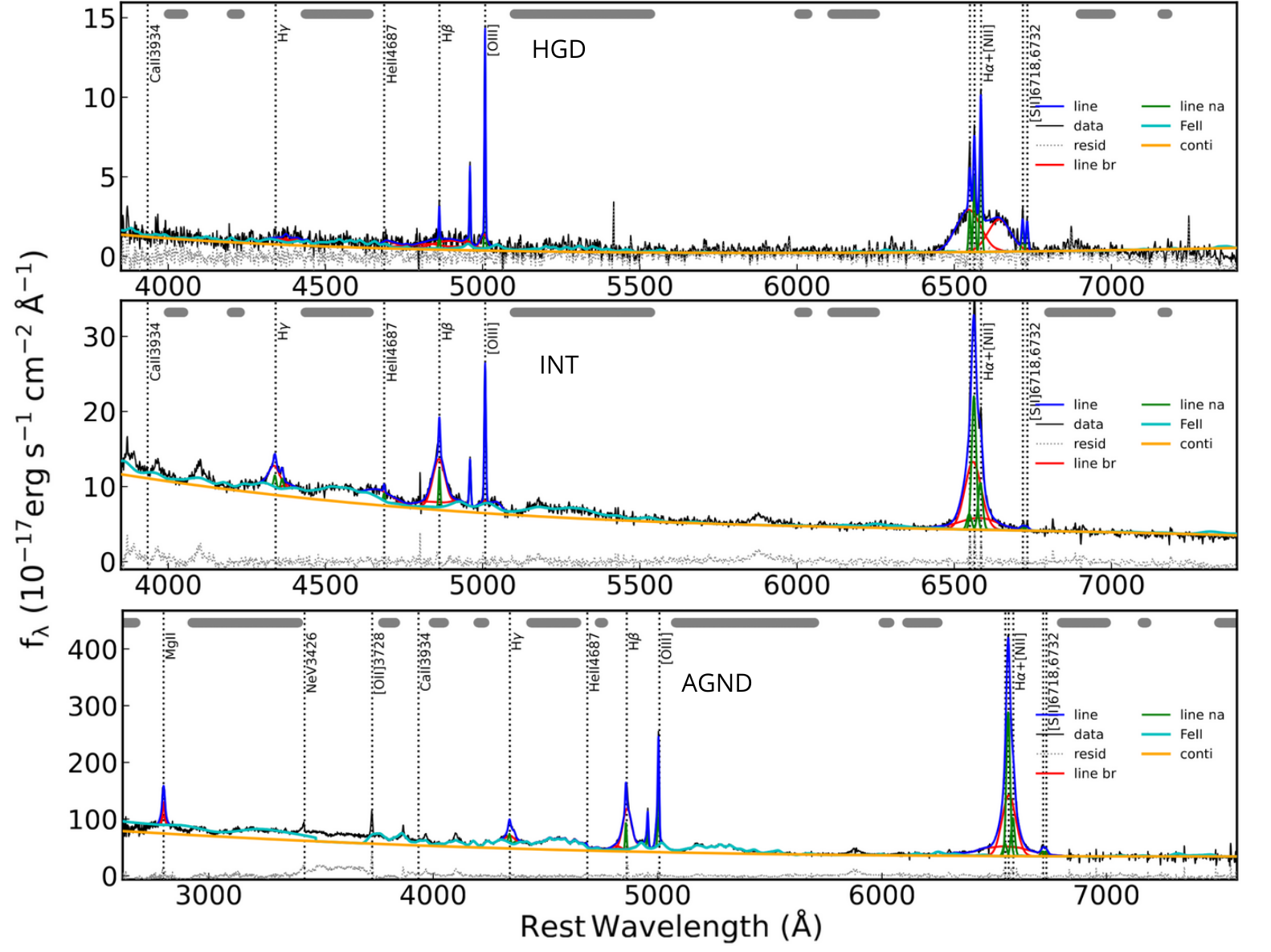}
    \caption{PyQSOFit fitting result example for the aforementioned groups. \textit{Top panel}: Host galaxy dominated example, \textit{medium panel}: Intermediate galaxy example, \textit{bottom panel}: AGN dominated example.}
    \label{fig:PyQSOfit}
\end{figure*}

We use the publicly available code PyQSOFit \citep{Guo2018PyQSOFit} 
to model the non-thermal underlying continua, along with the emission lines. The spectral decomposition includes the ranges of 
\hb\ and \ha\ regions for which we fitted the broad emission lines using Gaussian profiles, and modelled the \feii\ optical emission that could be particularly strong in the \hb\ vicinity. The code also includes Gaussian narrow emission lines and their wind components for the case of \oiiill. The fitting is done by a $\chi^2$-based method to fit the underlying continua, the spectral lines, and the \feii\ pseudo continuum template of \citet{borosongreen92} and \citet{VestergaardWilkes2001}. 
We note two features in the resulting stellar subtracted spectra from STL for some HGD objects. An absorption between 5140-5190 \AA, which appears to be a residual of the Mg absorption at 5200 \AA, and an emission in the range 6860-6920 \AA. 
These two features affect the underlying continuum placement, so we avoid those ranges in the spectral fitting of the HGD objects.

The QMS prescription uses Lorentzian and Gaussian profiles for Pop. A and B objects, respectively. However, as PyQSOFit does not natively support alternative line profiles to the Gaussian one, an independent fitting process was employed, utilizing the Voigt profile for the \hb\ line to standardize the profiles for the Eigenvector 1 analysis. To achieve this, we subtracted the PyQSOFit fitting for the continuum, \feii, and \oiiill\ emissions. The decision to exclude these components was made to prevent PyQSOFit from inadvertently removing the narrow component of \hb\ during the processing of the fitting results. 
Moreover, the primary reason for distinguishing between Gaussian and Voigt profiles was the fact that using two different profiles, such as Gaussian and Lorentzian, results in a clear discontinuity when deriving Eigenvector 1. The reason is that the FWHM of a Gaussian is approximately 1.18 times that of a Lorentzian if their dispersion parameters are equal. The FWHM difference is important in spectral decomposition of large samples because it could lead to a gap around 4000 \kms\ when the QMS is constructed. By employing a Voigt profile, this discontinuity is significantly reduced, leveraging the convolution of a Gaussian and a Lorentzian profile, improving the final results. After subtracting these components, the fitting was performed using a simple $\chi^2$ minimization algorithm to maintain consistency with the rest of the analyses.


\section{Results}
\label{sec:results}

\begin{figure}
    \includegraphics[scale=0.5]{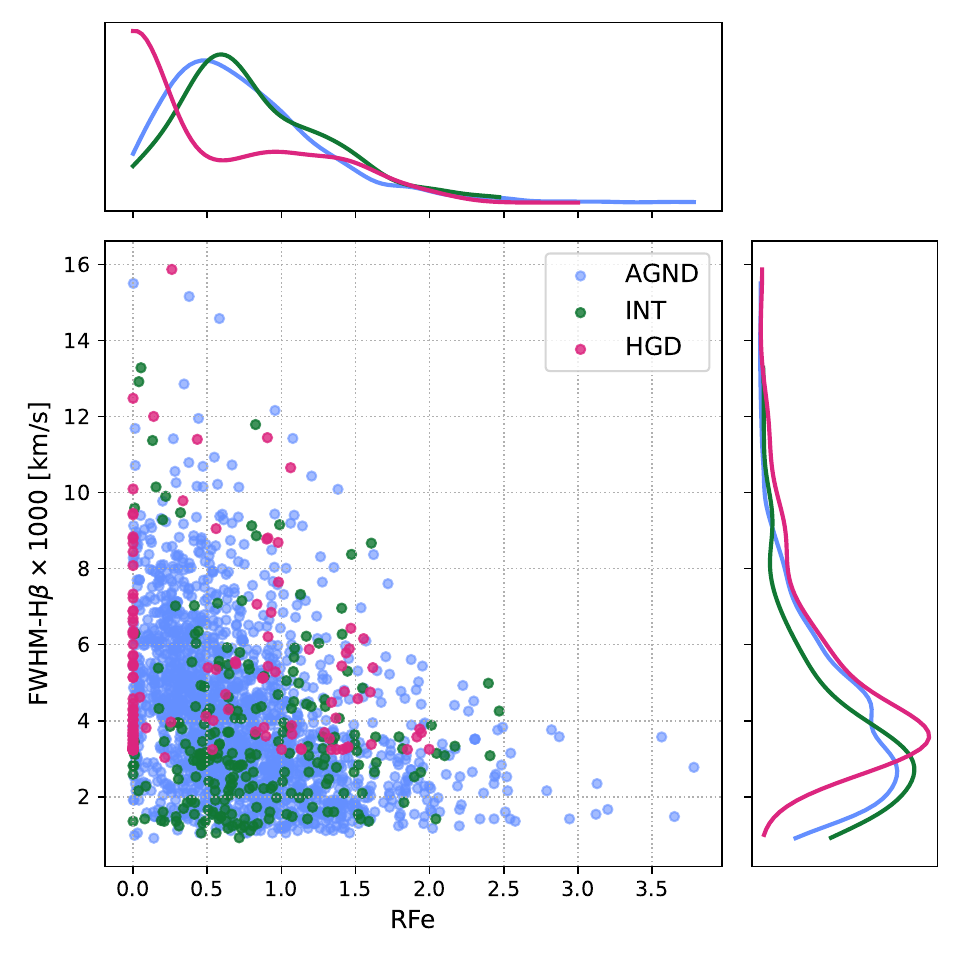}
    \includegraphics[scale=0.5]{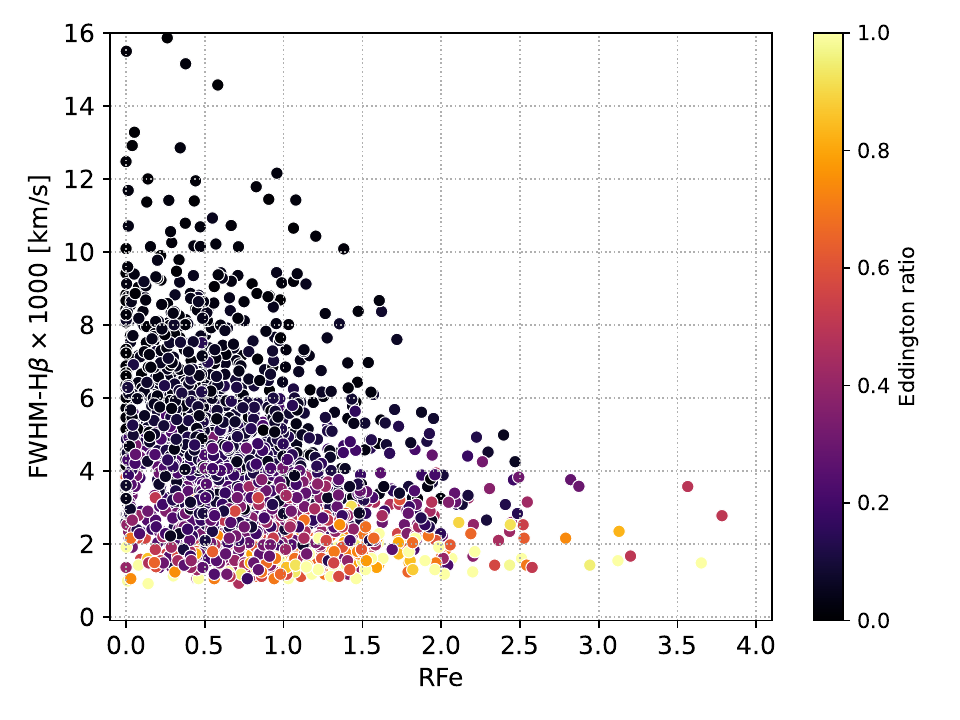}
    \includegraphics[scale=0.5]{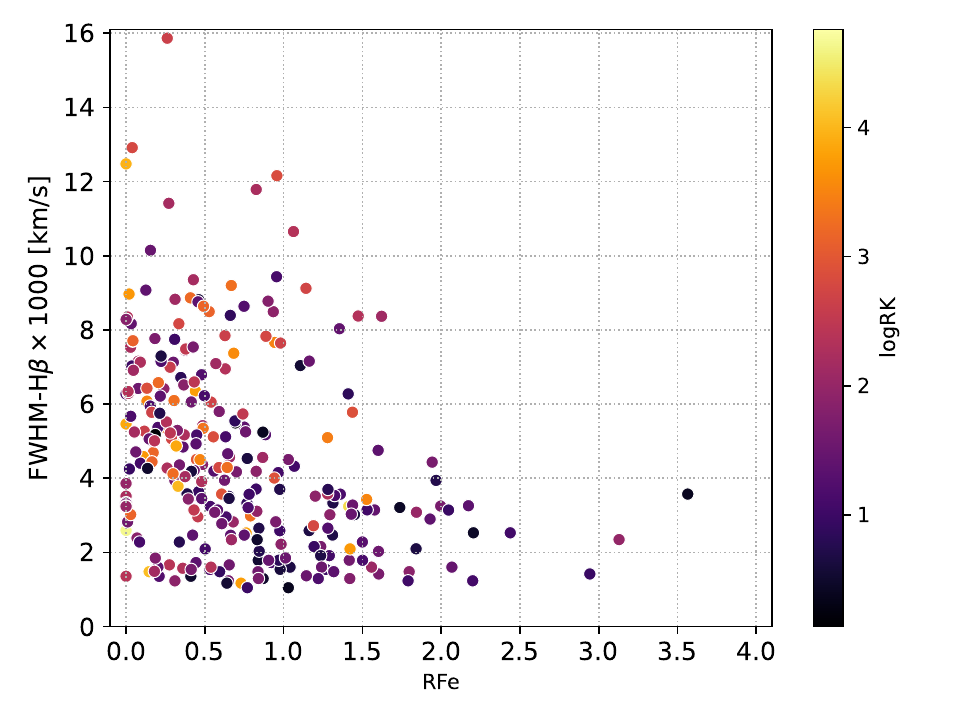}\\
    \caption{\textit{Upper panel}: Eigenvector 1. The x-axis measures the ratio between the equivalent width of the iron blend and \hb\, and the y-axis represents the FWHM of \hb. The colour code follows the previously mentioned scheme: blue for AGN, green for Intermediate objects, and red for HD. 
    \textit{Middle panel}: Quasar main sequence with estimated Eddington ratios in colour code. 
    \textit{Bottom panel}: Eigenvector 1 with colour code representing the log\rk' radio emission for the radio detected objects.
    }
    \label{fig:QMS}
\end{figure}

In this section, we will derive the optical QMS using the resultant spectral measurements, stressing the view of the AGN groups according to the AGN-HG contribution. 
After completing the spectral AGN modelling analysis, the FWHM of the \hb\ line and the equivalent widths of both \hb\ and \feii\ were used to derive the two main parameters of the QMS. This can be seen in Figure \ref{fig:QMS}. After fitting the lines, 
a cleaning process was carried out to identify and reanalyse outlier objects showing particular characteristics. These objects were re-examined to correct their positioning within the QMS, considering the following criteria. 

\begin{itemize}
    \item We detected several objects with very low \hb\ and/or \ha\ emissions and FWHM(\hb) $>$ 12000 \kms. So, as a first step to know the reliability of the detection, we established a lower limit for the height of the line to be larger than 1$\sigma$ of the adjacent RMS noise of a continuum window. We compute the height $h$ using the formula for the area $A=b\times h$, where $A$ is $flux$(line), the integrated flux for the line (\hb\ or \ha), and $b$ is the width of the base of the line (FWZI) in \AA. We assume that FWZI$\sim3\times$FWHM. In this way,
    \begin{equation}
        h_{\mathrm{line}} = \frac{c\text{flux(line)}}{3\lambda_{\mathrm{line}}\text{FWHM(line)}} 
    \end{equation}
    where $c$ is the speed of light and $\lambda_{line}$ is the rest frame wavelength of \ha\ or \hb. Objects with $h$ below the 1$\sigma$ criteria were excluded.
    
    \item We also detected objects with very low or practically no iron content, which the PyQSOFit fitting program overestimated, resulting in high \rfe\ values. These were corrected by visually inspecting the spectra and selecting objects that appeared as outliers (with \rfe\ $>$ 4) in the plot. For these objects, the ratio \rfe\ = 0. 
    Another feature for some HGD outliers with artificially large values of \rfe, is that not only the \feii\ is faint, but also their \hbbc\ is faint, with an underlying continuum somewhat underestimated. 
    These are some of the red dots of the HGD objects that have \rfe\ values between 1 and 2 in Figure \ref{fig:PyQSOfit}. For INT objects, there is a possibility that the \feii\ is overestimated due to HG subtraction or PL placement. However, by visual inspection of the spectral fits, we find that the effect is much smaller than in the case of HGD objects. 
    
    \item Part of the group of objects lacking the \hb\ line 
    show a strong \ha\ broad emission, 
    some of them showing profiles in agreement with an accretion disk model, i.e, double Gaussians as the one of the top panel of Fig. \ref{fig:PyQSOfit}. For these objects, the FWHM(\hb) value was estimated from the FWHM(\ha) line following 
    \cite{Shen:2007jt}. 
    The object selection was performed through visual inspection of a group of objects that appeared as outliers in the QMS with FWHM(\hb) values usually larger than 12,000 \kms.
    
    \item In the other part of objects 
    lacking the \hb\ line, 
    the \ha\ broad emission is also faint,
    with Gaussian-like profiles of FWHM $<$ 4,000 \kms, 
    i.e. the line profile is incompatible with the prescription of a Population A Lorentzian profile. We propose that the emission of this ``pseudo'' broad component is originating not from the Broad Line Region, but likely in the form of winds, as they also show broadened \oiiill. 
\end{itemize}


The histograms of the top panel of Figure \ref{fig:QMS} show the distribution of our sample in the QMS according to the HG-AGN dominance.
The HGD group represented as red dots has two distributions along the \rfe\ axis; one concordant with \rfe\ = 0 and another spread along \rfe\ values up to $\sim$two. The large majority (66\% of the HGD objects) have FWHM(\hb) $>$ 4000 \kms. 
We found 17\% of the HGD objects with \rfe\ $>$ 1 and FWHM(\hb) $>$ 4000 \kms. Looking at their spectra, we recognize the effect of the HG residuals of the stellar continuum subtraction in the \feii\ computations, making it difficult to establish the membership of the residual emission of the HG subtraction in the \feii\ emission region. 

The INT objects have a wide \rfe\ range, where most of the objects ($\sim$60\%) have \rfe\ values lower than one. The tail of INT quasars towards \rfe\ values larger than one has a similar explanation as for the HGD cases. The fitted \feii\ could be part of the stellar residuals from the STL modelling. 

Due to the fact that the AGND group is the most numerous, it is distributed throughout the whole QMS. The distribution follows previous QMS trends as seen, e.g., in \citet{zamfir10, shenho14, negrete18}, 
with FWHM(\hb) between 1000-16000 \kms, and \rfe\ values between 0-3.6. We note scattered objects for \rfe\ $>$ 2 which coincides with profiles called ``Peculiar'' by \citet{negrete18}, in the sense that they possess an extraordinary emission of \feii. The FWHM distribution seems to have a bimodality with two peaks at $\sim$3,000 and $\sim$5,000 \kms, with a tail up to 10,000 \kms\ and a few objects with FWHM up to 16,000 \kms

We also present in the middle panel of Figure \ref{fig:QMS} the trend of the \redd\ along the QMS. The \mbh\ used to compute the \ledd\ was estimated following \citet{ShenLiu2012} using the FWHM(\hbbc) of the Voight profiles and the continuum luminosity at 5100 \AA. The larger \redd\ values were found in objects with FWHM(\hbbc) $<$ 4000 \kms, with some objects reaching log\redd\ $\sim$ 1 in the \rfe\ $>$ 1 region. On the other hand, objects with broader FWHM have lower values of \redd. 



\section{The index diagram in a multiwavelength context}
\label{sec:discussion}
We use additional diagrams that identify AGN in Radio, soft X-ray and mid-infrared (MIR) ranges to revise the distribution of the AGND, INT and HGD classes. 

\subsection{Radio emission in the QMS and Index diagram} 

\begin{figure*}
    \centering
    \includegraphics[width=0.99\linewidth]{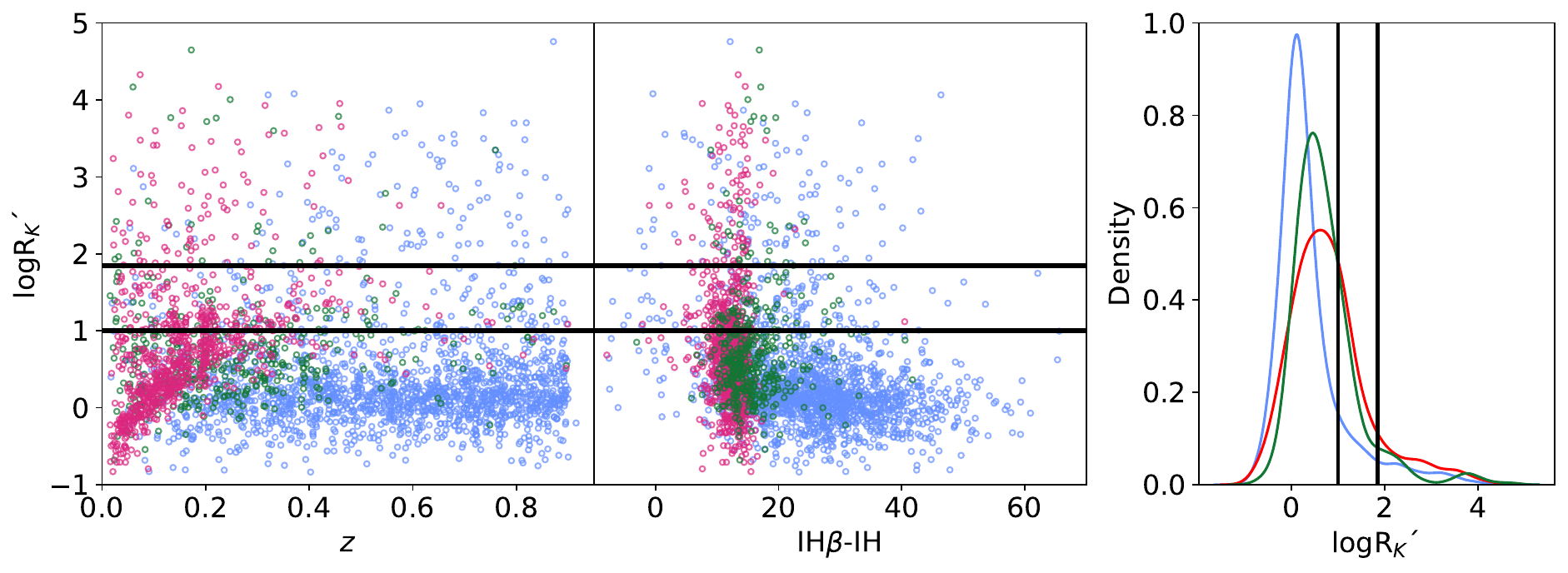}
    \caption{Distribution of log\rk' along the three AGND (blue), INT (green), and HGD (red) groups. Horizontal black lines in the left and middle panels, and vertical black lines in the left panel represent the limits for radio detected objects log\rk' $<$ 1, radio intermediate 1 $>$ log\rk' $>$ 1.85, and radio loud log\rk' $>$ 1.85, following \citet{ganci19}. }
    \label{fig:RK}
\end{figure*}

Radio emission of non-thermal origin is another fundamental characteristic of quasars. Most of this non-thermal radio emission is due to synchrotron emission, i.e., radiation by charged particles spinning at relativistic velocities through magnetic fields. 
This Section aims to visualize the radio emission within the QMS and the groups according to the AGN-HG contribution. We also include spectra with no \feii\ emission. 

The range of radio emission intensity is wide. However, it seems to be a clear dichotomy in this property. Historically, it has been proposed to separate the AGN depending on their ‘radio loudness’ through the parameter \rk, a measure of the ratio of radio (5 GHz) to optical (B-band) monochromatic luminosity \citep{Kellermann1989}. Radio Loud (RL) and Radio Quiet (RQ) objects were defined for AGN with \rk\ values greater or lower than 10, respectively. In large samples like the SDSS, about 10-15\%\ of all luminous AGN are radio loud \citep{UrryPadovani1995, Zamfir2008, Padovani2017AARv}.
More recently, with more sensitive radio detectors detecting faint flux densities, a wide population of star-forming galaxies and RQ AGN has changed the radio loudness denomination, transforming it into jetted and non-jetted objects \citep[][\citeyear{Padovani2017AARv}]{Padovani2016}. However, we will use the radio loudness nomenclature since faint flux radio data requires deep radio surveys and multi-frequency data \citep{Bonzini2012}.

In the QMS context, it has been shown that there are trends between Populations A and B with respect to the radio loudness of AGN \citep{sulentic00a, Sulentic2003, Zamfir2008, ganci19}. RL objects, including those with powerful jets, tend to be found predominantly in Population B, where \redd\ is low and \mbh\ is higher (\redd\ $<$ 0.2, mean value log\mbh\ = 8.9 for the FiSa). On the other hand, in Population A, where high \redd\ values are prevalent, even close to one, and \mbh\ are lower (\redd\ $>$ 0.2, mean value log\mbh\ = 7.9 for the FiSa), most objects are RQ with radio power compatible with emission by mechanisms of star formation processes, mainly for populations with \rfe\ $>$ 1 \citep{Sani2010, ganci19}. 

We cross-matched our CoSa with the objects reported in \citet{Stasinska2024}, which uses the Radio sources associated with Optical Galaxies and having Unresolved or Extended morphologies (ROGUE-I\footnote{\url{http://rogue.oa.uj.edu.pl/index.php?page=rogueI}}) 
catalogue. The ROGUE-I catalogue is a compilation of radio sources at $z <$ 0.64 obtained by cross-matching galaxies from the SDSS DR7 and radio sources from both the Faint Images of the Radio Sky at Twenty Centimetres (FIRST) survey and the NRAO (National Radio Astronomy Observatory) VLA (Very Large Array) Sky Survey (NVSS). We find 92 coincidences. We also searched our objects in \citet{BestHeckman2012}, who worked with AGN from the SDSS DR7 at $z <$ 0.3, finding 43 objects. 

In addition, we searched for radio data for our CoSa (limited to $z <$ 0.9) on catalogues from FIRST and NVSS. The FIRST Survey reports a catalogue detection limit at the source position of 0.98 mJy/beam (with a resolution $\theta$ = 5.4 arcsec), while for the NVSS, the detection limit is 2.3 mJy beam$^{-1}$ (with $\theta$ = 45 arcsec). 
We started to look for our CoSa objects (3259 objects) in FIRST, discarding objects outside the survey's field of coverage and objects that show a single point source emission off-centre, farther than 10 arcseconds from the reported optical position (375 objects). Once we have this filter selection, we perform a search with a radius of 50 arcseconds. For a large number of objects (2485), FIRST only reports the RMS at the search position that we used as an upper limit. For the rest of the objects, FIRST reported radio flux detections between 0.6 and 14600 mJy. We visually examined those images to corroborate that the radio emission belonged to the central galaxy, finding 314 objects having a single point source radio emission at the optical position, and 85 having resolved emission associated with the centre of the optical position.
On the other hand, from the NVSS search, we found 317 CoSa objects, for which 67 objects were not in the FIRST field of coverage, having values between 2.3 and 695.2 mJy, 119 objects have similar fluxes within $\pm$10\% of the FIRST flux, and 130 objects have fluxes three times higher on average than FIRST. The reported NVSS flux density values are expected to be systematically higher than FIRST, as the latter is not sensitive to extended sources like the former \citep{BestHeckman2012}.
The resulting search in \citet{Stasinska2024} and \citet[][92 and 43 objects respectively]{BestHeckman2012}, coincides with the values of the objects found in FIRST /NVSS, except for 12 high-z objects described below. We applied a Kolmogorov-Smirnov test to quantify the similarity within the value distributions of our estimations with the ones of \citet{Stasinska2024} and \citet{BestHeckman2012}, finding a p-value of 0.9 in both cases.

We used a modified Kellermann’s parameter \rk' with the $g$ band and the specific radio flux at 20 cm \citep[e.g.][]{ganci19}. We also identified and discarded 12 high $z$ (0.93 - 3.17) objects with log\rk' = 1.4 - 4.1 
that were selected when we crossmatched our sample with the \citet{Stasinska2024} and \citet{BestHeckman2012} lists, as they did not perform a correction for wrong $z$ estimation. We noticed that they are objects referred to in Section \ref{sec:z-corr}. 
We correct the flux of the $g$ band in the \rk' ratio by the K-correction for the HGD group. The HGD AGN are the most affected by the reddening caused by the redshift as they have the largest HG contribution. In contrast, INT and AGND samples, which are dominated by the contribution of the AGN, are not affected because their spectrum shows a constant slope across their wavelengths. When these objects are shifted, they will not show any reddening. The colour correction was performed considering the shift for the photometric $g$ band towards bands $r, i$, and $z$. However, the HGD AGN \rk' can be considered as an upper limit due to the dominance of the HG in the $g$ band. 
Lower panel of Figure \ref{fig:QMS} presents the log\rk' values in the QMS for objects with radio detections, while Figure \ref{fig:RK} shows the distribution of log\rk' along the three AGN groups defined in Sec. \ref{sec:IndexDiagram}, AGND, INT, and HGD for the FiSa objects (including the RMS upper limits). 
Left panel presents the \rk' distribution with respect to $z$, 
The distribution of the difference of \hb\ and H indices with respect to \rk' is given in the middle panel. The histogram of the right panel shows the distribution of the \rk' values, where we find a tendency of the median values according to the AGN-HG dominance: 0.17 for AGND, 0.57 for INT, and 0.68 for HGD objects. 


\begin{figure*}
    \centering
    \includegraphics[width=0.99\linewidth]{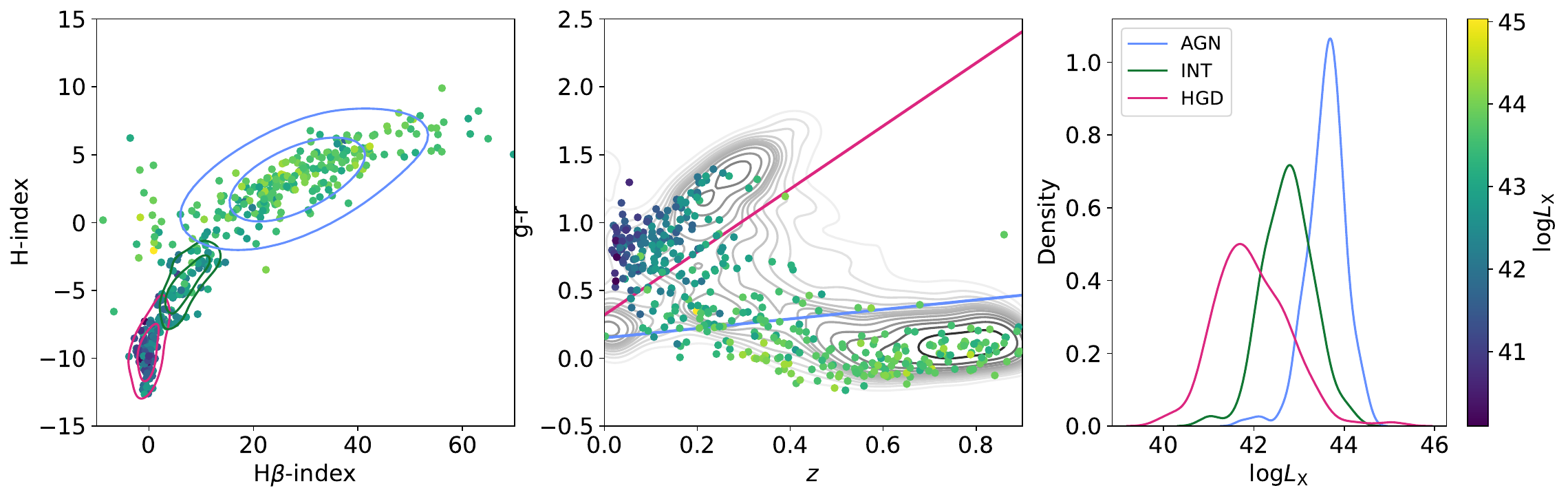}  
    \caption{Distribution of the 0.2-2.3 keV X-ray eROSITA luminosities within the H-\hb\ index diagram (Left panel), Colour--$z$ diagram, and their distribution in function of the AGN, INT and HD classifications. The red, green and blue contours/lines represent the HGD, INT and AGN regions and distributions.}
    \label{fig:rX}
\end{figure*}

\subsection{X-ray fluxes}
On the other hand, one of the consequences of the accretion of matter into SMBHs is the emission of X-rays \citep{Merloni+2012}. Therefore, studying the X-ray luminosities provides an alternative way to classify the AGN contribution \citep[][]{Fiore+2003, Fiore+2012, Aydar2025arXiv}. We make use of the soft X-ray fluxes (0.2-2.3 keV) from the extended ROentgen Survey with an Imaging Telescope Array \citep[eROSITA,][]{Predehl+2021, Merloni+2024} taken from the SDSS-V/eROSITA Data Level 1 (DL1) from the eROSITA Final Equatorial Depth Survey (eFEDS) catalogue \citep[eFEDS][]{Brunner+2022, Aydar2025arXiv}. 

We show in Figure \ref{fig:rX} the colour-$z$ and H-H$\beta$ index diagrams of Figure \ref{fig:color-index}, but now showing the distribution of their X-ray SDSS-V/eROSITA DL1 luminosities and the distribution of those luminosities for our three galaxy classifications: AGN, INT and HGD. As expected, the AGN-dominated spectra have X-ray luminosities ($L_X$ in \ergs) between log$L_X$ = 43-44 dex 
with a median of 43.63 dex, while the intermediate spectra have an X-ray luminosity with a median of 42.79 dex within a range between 42 to 43.5 dex. 
Finally, the host dominated spectra have a median of log$L_X$ = 41.86 dex and a range between 41 to 43 dex. 
As it is discussed in \citet{Aydar2025arXiv}, these results indicate that most objects showing host-dominated spectra, that are not necessarily active galaxies, have soft X-ray luminosities below log$L_X$ = 42 
\citep{Nandra+2002, Brandt+2005}, while most of the AGN dominated spectra have X-ray luminosities larger than log$L_X$ = 43.5. 
The AGN dominated objects are the galaxies with the most intense nuclear activity, and therefore, we can define a lower limit of log$L_X$ = 43.5 dex 
to identify those objects. Finally, we can define the range between log$L_X$ = 42 to 43.5 dex 
to define the region where the intermediate AGN spectra can lie, and as it is mentioned in \citet{Aydar2025arXiv}, this sample can have a large diversity in their optical-UV emission line properties.

\subsection{Infrared diagrams}

\begin{figure}
    \centering 
    \includegraphics[width=0.99\linewidth]{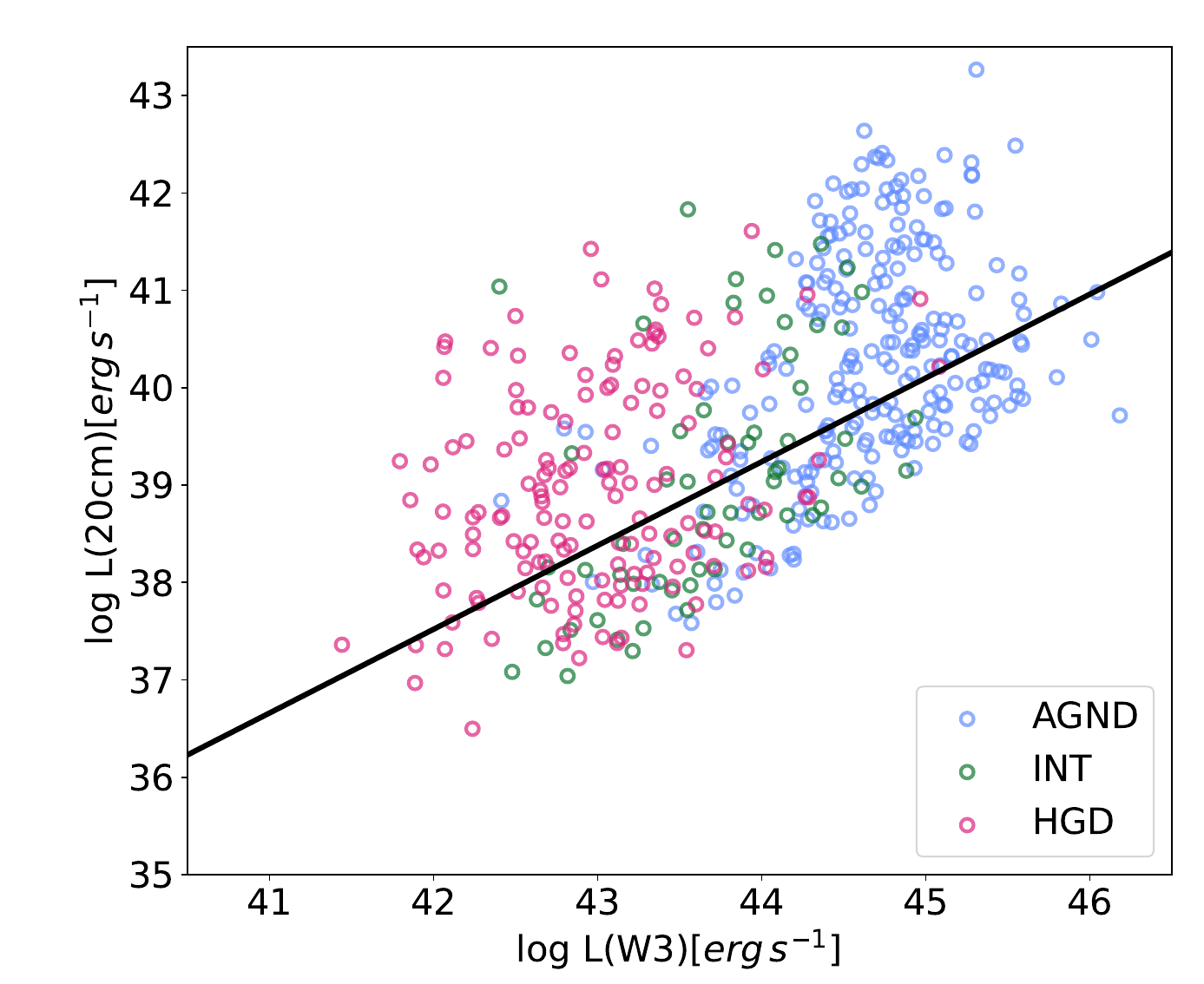}
    \includegraphics[width=0.99\linewidth]{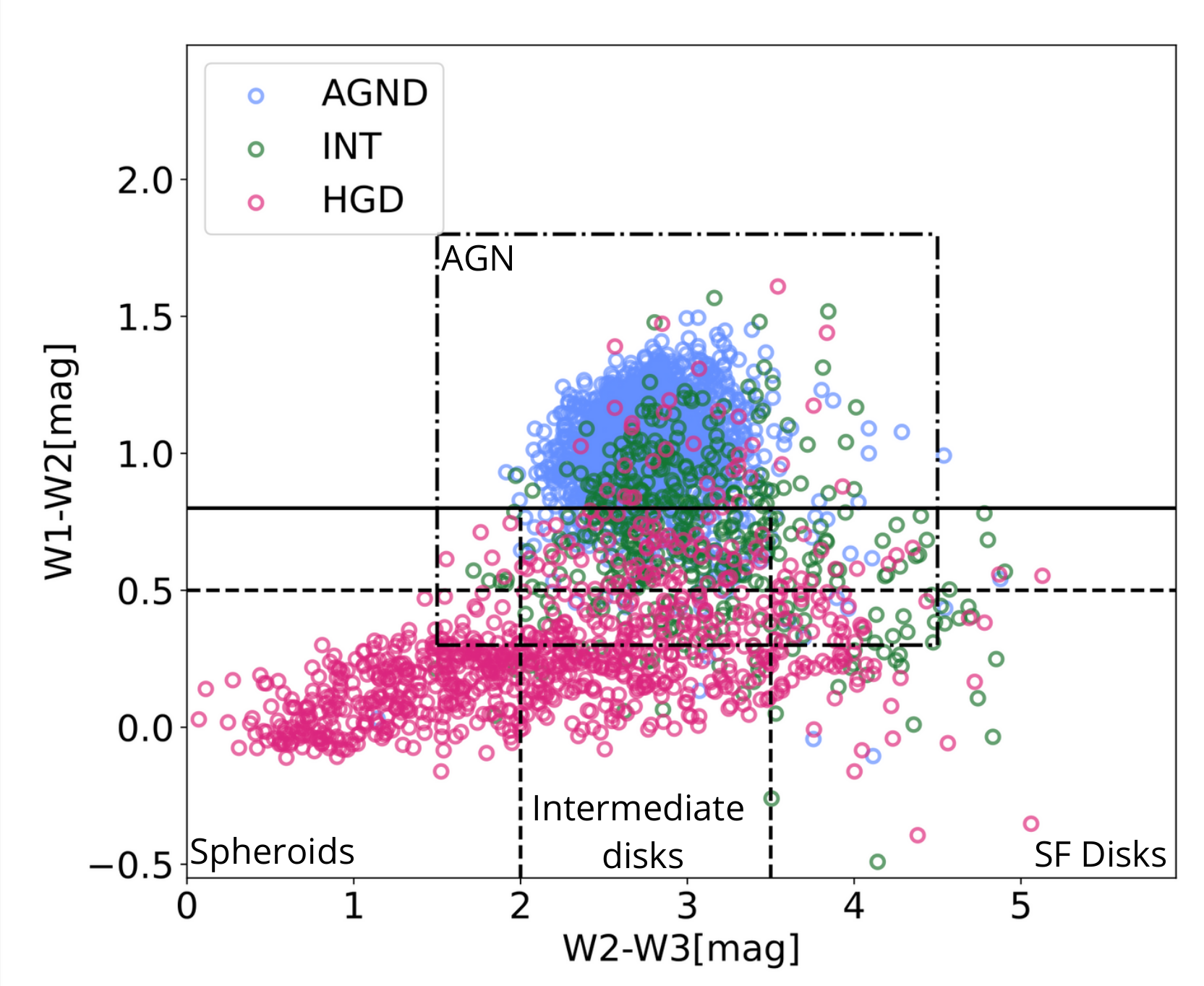}
    \caption{\textit{(Upper panel)} AGN-HG distributions of the radio/mid-IR star formation correlation of \citet{Mingo2016} considering objects with radio detections. Above the solid line, the radio emission is considered to arise from an AGN. \textit{(Lower panel)} Color WISE diagram of \citet{Jarrett2017} that delimits regions of galaxies.}
    \label{fig:wise-radio-colors}
\end{figure}

For completeness in the multi-wavelength AGN-HG dominance analysis, we also search for information of the CoSa (at $z <$ 0.9) in the infrared (IR) using the Wide-field Infrared Survey Explorer (WISE) database. 
Using also the radio data retrieved above, we considered the radio/mid-IR star formation correlation to set the origin of the radio emission, especially for the radio-detected and intermediate-radio objects. We employed the limit proposed by \citet{Mingo2016} (log(L$_{20cm}$) = (0.86 $\pm$ 0.04) log(L$_{12\mu m}$) + (1.4 $\pm$ 1.5)) in the correlation between the radio luminosity ($L(20cm)$) and the infrared luminosity at 12$\mu m$ (WISE W3 filter $L(W3)$), which is shown in the upper panel of Figure \ref{fig:wise-radio-colors}. Objects above this limit are classified as AGN radio sources; below this limit, the radio emission is considered to arise from star formation regions. 
We also see a bimodality of the HGD and AGND distributions, with the INT objects in between. The AGN-HG dominance distribution coincides with the separation into elliptical, spiral, and AGN objects of \citet[][see their Fig. 13]{Mingo2016}. HGD objects fall into the elliptical galaxies region, with some objects in the spiral region. INT objects coincide with the spirals and AGN region, while AGND objects are in the AGN region. This separation is also stressed in the IR color-color diagram of the lower panel of Fig. \ref{fig:wise-radio-colors}, proposed by \citet[][see their Fig. 11]{Jarrett2017}, where the regions for ellipticals (spheroids), spirals (intermediate disks), and star-forming disks are delimited according to the WISE colour regions. 

\begin{table}
	\centering
	\caption{Median parameters according to AGN-HG dominance.}
	\label{tab:Parameters}
	\begin{tabular}{lcccccccc} 
		\hline
		Parameter & HGD & INT & AGND & Sample \\
		\hline
		$g-r$ & $0.81 \pm 0.01$ & $0.45 \pm 0.02$ & $0.03 \pm 0.00$ & CoSa\\
        $z$ & $0.10 \pm 0.01$ & $0.25 \pm 0.01$ & $0.69 \pm 0.02$ & CoSa\\
        H-index & $-7.77 \pm 0.22$ & $-4.40 \pm 0.01$ & $2.64 \pm 0.02$ & CoSa\\
        \hb-index & $0.73\pm 0.20$ & $9.57 \pm 0.61$ & $26.12 \pm 0.11$ & CoSa\\
        \redd & $0.03 \pm 0.01$ & $0.02 \pm 0.01$ & $0.52 \pm 0.27$ & FiSa\\
        \mbh & $7.91 \pm 0.04$ & $7.84 \pm 0.05$ & $8.61 \pm 0.03$ & FiSa\\
        log\rk & 0.78 $\pm$ 0.87 & 0.67 $\pm$ 0.75 & 0.35 $\pm$ 0.73 & FiSa\\ 
        logL$_X$  & 41.86 $\pm$ 1.08 & 42.79 $\pm$ 0.87 & 43.63 $\pm$ 0.45  & FiSa\\
        logW1-W2  & 0.54 $\pm$ 0.37 & 0.75 $\pm$ 0.27 & 1.05 $\pm$ 0.16  & CoSa\\
        \hline
	\end{tabular}
\end{table}

\section{Summary}
\label{sec:summary}

We use the spectroscopic data from phase V of the SDSS, especially the data products from the Black Hole Mapper, comprising 177,977 spectra that were first filtered using colour criteria and S/N ratios (S/N $> 20$). Then, a methodology similar to \citetalias{CortesSuarez2022} was applied to obtain candidates to be Type 1 AGN objects using the flux ratios between regions around \ha\ and the continuum around 6410 \AA. An analogous method was also applied to \hb\ comparing with the continuum around 4,960 \AA\ in order to identify candidate objects for the determination. An additional $z$ correction was applied to the CoSa objects, as $\sim$140 of the objects had a wrong $z$ assignment. 

The selected sample was then processed using the colour-$z$ diagram to identify 
two main regions: reddened objects at low redshift ($z <$ 0.5) dominated by the stellar continua, and blue objects at higher redshifts with a transition region between the regions. Then, the H-\hb\ index diagram further refines the classification considering both the stellar absorptions and AGN emission lines 
revealing trends in the dominance of the HG or the AGN. The index diagram leads us to define three groups: the host galaxy dominated (HGD), the AGN dominated (AGND), and intermediate (INT) spectra. 
The selection of the objects between the two different methods is consistent with each other. 

For the spectral line modelling, we employed two steps. Once the sample is divided by the HG-AGN dominance, 
we use the Starlight code to perform a stellar emission decomposition in both the INT and the HGD objects. After removing the HG contribution, we use the PyQSOFit code to model the AGN emission line spectra for the three groups. 
Finally, as the code does not support other profiles apart from the Gaussian, we included a Voigt profile to carry on with the QMS analysis. 

We built the QMS, emphasising the location of the three groups. HGD objects tend to show FWHM(\hbbc) $>$ 4000 \kms, and no \feii\ emission. Some outliers towards \rfe\ values $>$ 1 were found resulting from the over-fitting of \feii\ due to the residuals of the HG subtraction, and the low contribution of \hbbc. INT and AGND objects were found with a wide distribution of \rfe, with some INT objects influenced by subtraction residuals from the stellar component. The AGND spectra showed a complete distribution along the QMS, following known patterns with high FWHM and \rfe\ values.

Radio, X-ray, and infrared emissions were analysed to better understand the relative contribution of AGN-HG dominance. 
We obtained radio data for 12.2\% of the CoSa. For another 75.2\% of the objects, we considered the RMS at the search position as an upper limit. Using a modified Kellermann's parameter \rk', we find an increasing tendency of the median values of \rk' towards the HGD objects, with median values of log\rk' equal to 0.17 for AGND, 0.57 for INT, and 0.68 for HGD objects. On the other hand, we determine a decreasing tendency of the soft X-ray luminosities from SDSS-V/eROSITA DL1, towards the HGD objects, having median values of the X-ray luminosities of logL$_X$ equal to 43.63 for AGND, 42.79 for INT, and 41.86 for HGD objects. 

MIR data were obtained to distinguish whether the radio emission comes from the AGN or from star-forming regions, as well as to delimit galaxy regions. In the correlation between radio and IR luminosity, a bimodality in the distribution of HGD and AGND objects was observed, with INT objects falling in between. This distribution coincides with the morphological classification of galaxies (ellipticals, spirals and AGN), and is confirmed by a color-color diagram in the IR, where each type of object occupies well-defined regions according to its spectral properties. Table \ref{tab:Parameters} summarizes the median of all the parameters discussed throughout the paper, according to AGN-HG dominance.

\section*{Acknowledgements}
CAN and HIM acknowledge the support from grants SECIHTI CBF2023-2024-1418 and CF-2023-G-543, PAPIIT UNAM IA104325, IN-106823, and IN-119123. SFS acknowledges support by the PASPA program of the DGAPA-UNAM. RJA was supported by FONDECYT grant number 1231718 and by the ANID BASAL project FB210003. PRH acknowledges support by the Sloan Digital Sky Survey's FAST 4ward program, through the National Science Foundation award AST-2425222. MLM-A acknowledges financial support from Millenium Nucleus NCN2023${\_}$002 (TITANs), ANID Millennium Science Initiative (AIM23-0001), and the China-Chile Joint Research Fund (CCJRF2310).

Funding for the Sloan Digital Sky Survey V has been provided by the Alfred P. Sloan Foundation, the Heising-Simons Foundation, the National Science Foundation, and the Participating Institutions. SDSS acknowledges support and resources from the Center for High-Performance Computing at the University of Utah. SDSS telescopes are located at Apache Point Observatory, funded by the Astrophysical Research Consortium and operated by New Mexico State University, and at Las Campanas Observatory, operated by the Carnegie Institution for Science. The SDSS website is \url{www.sdss.org}.

SDSS is managed by the Astrophysical Research Consortium for the Participating Institutions of the SDSS Collaboration, including Caltech, The Carnegie Institution for Science, Chilean National Time Allocation Committee (CNTAC) ratified researchers, The Flatiron Institute, the Gotham Participation Group, Harvard University, Heidelberg University, The Johns Hopkins University, L'Ecole polytechnique fédérale de Lausanne (EPFL), Leibniz-Institut für Astrophysik Potsdam (AIP), Max-Planck-Institut für Astronomie (MPIA Heidelberg), MaxPlanck-Institut für Extraterrestrische Physik (MPE), Nanjing University, National Astronomical Observatories of China (NAOC), New Mexico State University, The Ohio State University, Pennsylvania State University, Smithsonian Astrophysical Observatory, Space Telescope Science Institute (STScI), the Stellar Astrophysics Participation Group, Universidad Nacional Autónoma de México, University of Arizona, University of Colorado Boulder, University of Illinois at Urbana-Champaign, University of Toronto, University of Utah, University of Virginia, Yale University, and Yunnan University.

\section*{Data Availability}

The data used in this paper will be publicly available in the forthcoming SDSS Data Release 19. 



\bibliographystyle{mnras}
\bibliography{example} 

\bsp	
\label{lastpage}
\end{document}